\pdfoutput=1

\documentclass[preprint,fleqn,5p,authoryear]{elsarticle}

\usepackage{graphicx}
\usepackage{amssymb,amsmath}
\usepackage{natbib}
\usepackage{hyperref}
\hypersetup{pdftitle = The title of my PDF, pdfauthor = My name, pdfsubject= The subject, pdfkeywords = keyword1 keyword2 keyword3}
\hypersetup{colorlinks = true, linkcolor = blue, anchorcolor = red, citecolor = blue, filecolor = red, pagecolor = red, urlcolor = blue}

\journal{Planetary and Space Science}

\begin{document}

\begin{frontmatter}

\title{Near-optimal capture in the planar circular restricted Pluto-Charon system}

\author[eez]{Euaggelos E. Zotos\corref{cor1}}
\ead{evzotos@physics.auth.gr}

\author[yq]{Yi Qi}

\cortext[cor1]{Corresponding author}

\address[eez]{Department of Physics, School of Science,
Aristotle University of Thessaloniki, GR-541 24, Thessaloniki, Greece}

\address[yq]{Department of Aerospace Engineering, Ryerson University,
350 Victoria Street, Toronto, ON M5B 2K3, Canada}

\begin{abstract}
In this paper, the near-optimal capture problem is numerically investigated to locate the optimal capture points around Charon using the Pluto-Charon planar circular restricted three-body problem (PCRTBP). Capture orbits are divided into the pre- and post-maneuver portions. In the pre-maneuver portion, the gravitational capture conditions are discussed by backward numerical integration. In the post-maneuver portion, the smaller alignment index (SALI) is applied to numerically investigate the orbital characters of long-term capture orbits around Charon. The initial conditions corresponding to three types of motion: (i) bounded (regular or chaotic), (ii) escaping and (iii) collisional are classified and studied. Combining results of the pre- and post-maneuver portions, the near-optimal capture method is presented to find optimal capture points by overlay figures, and then the corresponding capture orbits are constructed in the PCRTBP. The different results between the Pluto-Charon system and the Earth-Moon system are compared and analyzed. Our results obtained in this paper, including the optimal capture points and the corresponding capture orbits, could be applied in future space mission design.
\end{abstract}

\begin{keyword}
methods: numerical -- celestial mechanics -- planets and satellites: individual: Charon -- planets and satellites: individual: Pluto
\end{keyword}
\end{frontmatter}

\section{Introduction}
\label{intro}

Capture dynamics is a hot topic in the astronomy and astronautics. Using Kolmogorov-Arnold-Moser (KAM) tori, \cite{ABW03} proposed the chaos-assisted capture (CAC) mechanism in the circular restricted three-body problem (CRTBP) to investigate irregular moons of Jupiter and Saturn, which identifies chaos as the essential feature responsible for initial temporary gravitational trapping within a planet's Hill sphere. Their theory can explain why Saturn has more prograde irregular moons than Jupiter. In the Earth-Moon system, asteroid 2006 RH$_{120}$ was observed to be captured temporarily from September 2006 to June 2007. \cite{GVJ12} first proposed a concept of temporary capture (TC) to describe and investigate this phenomenon in the Earth-Moon system. Based on their computation, at any given time there should be at least one temporarily-captured irregular natural Earth satellite (NES) of 1-m diameter orbiting the Earth. The well-known gravitational capture or weak stability boundary (WSB) theory was first proposed by \cite{B87} to solve the lunar capture problem in Earth-to-Moon transfers. Then, this theory has been successfully applied to many lunar missions, such as JAXA's Hiten rescue \citep{BM90}, ESA's SMART-1 \citep{SHP01}, and NASA's GRAIL \citep{RF10}. The in-depth research indicates that the WSB can be extended to interplanetary transfers, such as to Mercury \citep{JCG04} and Mars \citep{TB15}.

In addition, many scholars focus on the capture problem of the near-Earth asteroids (NEAs), because the NEAs play important roles in the origin of the solar system and the establishment of early warning of asteroid impact. \cite{BCL10} proposed a method to search for low-energy NEAs that may be temporarily captured by the Earth. NASA's Asteroid Redirect Robotic Mission (ARRM) was proposed to capture an entire $4\sim 10$ m NEA and redirect it into an orbit around the Moon where astronauts in the Orion spacecraft could explore it \citep{SLMc13}. \cite{YSMc13} found a family of so-called Easily Retrievable Objects (EROs), which can be transported from accessible heliocentric orbits into the Earth's neighborhood at affordable costs. \cite{BYB15} applied lunar flybys and Earth flybys to capture NEAs into bounded orbits of the Earth. Based on the CAC mechanism in \citep{ABW03}, \cite{VMc14} studied the low-energy capture of asteroids onto KAM tori in the Sun-Earth system. \cite{Qd18} proposed a concept called short-term capture (STC) to investigate temporary capture phenomena of NEAs in the EMS. Space and time conditions of STC were defined and studied. Numerical results indicated that there was a clear association between the distributions of the time probability of STC and its space conditions. \par

The impulsive maneuver, $\Delta V$, provided by the propulsion system of the spacecraft is the most practical maneuver method in the space mission to the present day. Hence, the optimal capture strategy usually can be evaluated by the magnitude of the impulsive maneuver. \cite{QX14} proposed the near-optimal lunar capture theory in the Earth-Moon planar CRTBP to locate the optimal capture point and achieve the permanent lunar capture using the minimum $\Delta V$. This method combined the lunar gravitational capture and KAM tori around the Moon. The minimum capture velocity and the minimum capture eccentricity before the maneuver were defined and analysed. The permanent lunar capture after the maneuver was investigated by the distribution of KAM tori around the Moon. Furthermore, based on the near-optimal lunar capture, \cite{QX16} constructed low-energy transfers from the Earth to permanent lunar orbits. The numerical results indicated that compared with traditional patched-conic transfers, those low-energy transfers can efficiently save fuel costs, which actually were very close to the theoretical minimum. \par

In this paper, we apply the near-optimal capture theory to investigate the capture problem of Charon in the Pluto-Charon planar CRTBP. There are two main differences between this paper and the previous research of the Earth-Moon system in \citep{QX14}. Firstly, the mass ratio $\mu$ of the Pluto-Charon system is significantly larger than that of the Earth-Moon system, so the gravity of Charon is more significant in its region than the case of the Moon. Consequently, we postulate that the near-optimal capture in the Pluto-Charon planar CRTBP can show different characters and results. Secondly, \cite{QX14} applied KAM tori to analyze the long-term orbital characters after the impulsive maneuver $\Delta V$. Although this method is simple and visible, the position and coverage of KAM tori are based on the empirical observation rather than the quantitative analysis. In this paper, the smaller alignment index (SALI), a common index for determining the regular or chaotic character of an orbit, is applied to investigate and classify the long-term orbits around Charon more accurately instead of KAM tori.

In \cite{Z15} the orbital dynamics in the planar circular Pluto-Charon system were numerically investigated. More specifically, large sets of initial conditions of orbits were classified in an attempt to determine the dynamics of a test particle (e.g., a spacecraft, or a comet, or an asteroid) moving in the vicinity of Charon. It was found that the configuration $(x,y)$ space is covered by a highly complicated mixture of several types of basins, such as basins of escape, basins of collision and basins of bounded motion. In addition, by varying the value of the Jacobi constant (or in other words the energy level) he managed to monitor the parametric evolution of all types of orbits. In this paper, the quantitative method similar to \citep{Z15} will be proposed to analyze the long-term orbital dynamics in the Pluto-Charon system. Based on \cite{QX14}, the minimum $\Delta V$ satisfying the permanent capture can be obtained from two perspectives: decreasing the pre-maneuver velocity $ V_1$ and increasing the post-maneuver velocity $ V_2$. Under given constraints of the capture point and Jacobi constant, there exist a minimum $ V_1$ and a maximum $ V_2$. Combining two aspects, the optimal capture point with the minimum $\Delta V = V_1 - V_2$ can be found.\par

According to the above discussion, the structure of this paper is separated into five parts. Section \ref{mod} introduces the main properties of the dynamical system. Section \ref{prem} focuses on the gravitational capture conditions in pre-maneuver portion and obtain the minimum $V_1$. In Section \ref{postm}, we investigate the long-term capture in post-maneuver potion and obtain the maximum $V_2$. After that, the near-optimal capture of Charon in the Pluto-Charon planar CRTBP is proposed and analyzed in Section \ref{nearo}. Finally, conclusions of our investigation are presented in Section \ref{conc}.

\section{Description of the dynamical model}
\label{mod}

\begin{figure}[!t]
\centering
\resizebox{\hsize}{!}{\includegraphics{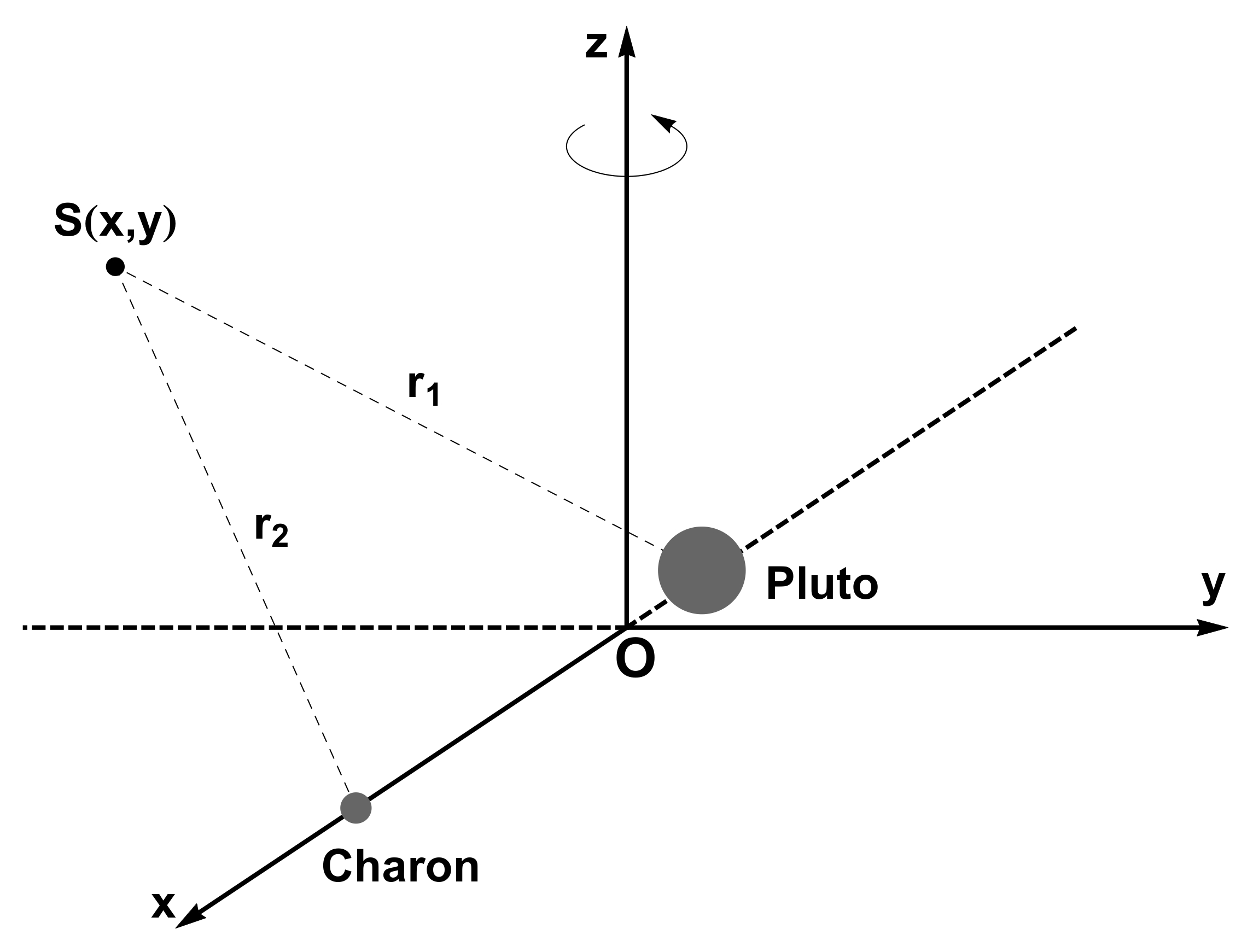}}
\caption{The planar configuration of the circular restricted Pluto-Charon system.}
\label{scheme}
\end{figure}

According to the theory of the classical circular restricted three-body problem \citep{S67}, the two primary bodies $P_1$ (Pluto) and $P_2$ (Charon) move on circular orbits \citep{BTG12}, with the same angular velocity, around their common center of mass. The third body, also known as the test particle (e.g., a spacecraft, or a comet, or an asteroid), moves on the same plane under the resultant Newtonian gravitational field of the two main bodies. Taking into account that the mass of the test particle is extremely small, compared to the masses of the two primaries, we can reasonably assume that the third body does not perturb the motion of the primaries. The mass ratio $\mu = m_2/(m_1 + m_2)$, with $m_1 > m_2$ defines the non-dimensional masses of the two primaries $m_1 = 1-\mu$ and $m_2 = \mu$. In this notation, the centers of Pluto and Charon are located at $(-\mu, 0)$ and $(1-\mu,0)$, respectively (see Fig. \ref{scheme}).

The planar motion of a test particle can be described by the following set of differential equations
\begin{align}
U_x(x,y) &= \frac{\partial U}{\partial x} = \ddot{x} - 2\dot{y}, \nonumber\\
U_y(x,y) &= \frac{\partial U}{\partial y} = \ddot{y} + 2\dot{x},
\label{eqmot}
\end{align}
where
\begin{equation}
U(x,y) = \frac{m_1}{r_1} + \frac{m_2}{r_2} + \frac{\mu\left(1 - \mu\right)}{2} + \frac{1}{2}\left(x^2 + y^2 \right),
\label{pot}
\end{equation}
is the time-independent effective potential function of the system, while
\begin{align}
r_1 &= \sqrt{\left(x + \mu\right)^2 + y^2}, \nonumber\\
r_2 &= \sqrt{\left(x + \mu - 1\right)^2 + y^2},
\label{dist}
\end{align}
are the distances to the respective primaries.

Similarly, the set of the variational equations is given by
\begin{align}
\dot{(\delta x)} &= \delta \dot{x}, \nonumber\\
\dot{(\delta y)} &= \delta \dot{y}, \nonumber\\
\dot{(\delta z)} &= \delta \dot{z}, \nonumber\\
(\dot{\delta \dot{x}}) &= \frac{\partial^2 U}{\partial x^2}\delta x + \frac{\partial^2 U}{\partial x \partial
y}\delta y + \frac{\partial^2 U}{\partial x \partial z}\delta z + 2 \delta \dot{y}, \nonumber \\
(\dot{\delta \dot{y}}) &= \frac{\partial^2 U}{\partial y \partial x}\delta x + \frac{\partial^2 U}{\partial
y^2}\delta y + \frac{\partial^2 U}{\partial y \partial z}\delta z - 2 \delta \dot{x}, \nonumber\\
(\dot{\delta \dot{z}}) &= \frac{\partial^2 U}{\partial z \partial x}\delta x + \frac{\partial^2 U}{\partial z
\partial y}\delta y + \frac{\partial^2 U}{\partial z^2}\delta z.
\label{variac}
\end{align}

The set of the equations of motion (\ref{eqmot}) admits only the following integral of motion
\begin{equation}
J(x,y,\dot{x},\dot{y}) = 2U(x,y) - \left(\dot{x}^2 + \dot{y}^2\right) = C,
\label{ham}
\end{equation}
which is also known as the Jacobi integral. The numerical value $C$ of the Hamiltonian remains constant and it is called the Jacobi constant.

According to \citet{url} the value of the mass ratio for the Pluto-Charon system is $\mu$ = 0.10851122058\footnote{It should be noted that in \citet{Z15}, the value $\mu = 0.099876695437731$ was adopted. However, over the last years the official value of the mass ratio regarding the Pluto-Charon planetary system has been revised.}, which is significantly larger than that of the Earth-Moon planetary system, 0.0121506683 \citep{QX14}. The values of the Jacobi constant at the five Lagrange points $(L_i, i = 1,...,5)$ are denoted by $C_i$, $i = 1,..., 5$ and they are equal to: $C_1$ = 3.717080, $C_2$ = 3.576079, $C_3$ = 3.204728, and $C_4 = C_5$ = 3.

\section{Gravitational capture}
\label{prem}

\begin{figure}[!t]
\centering
\resizebox{\hsize}{!}{\includegraphics{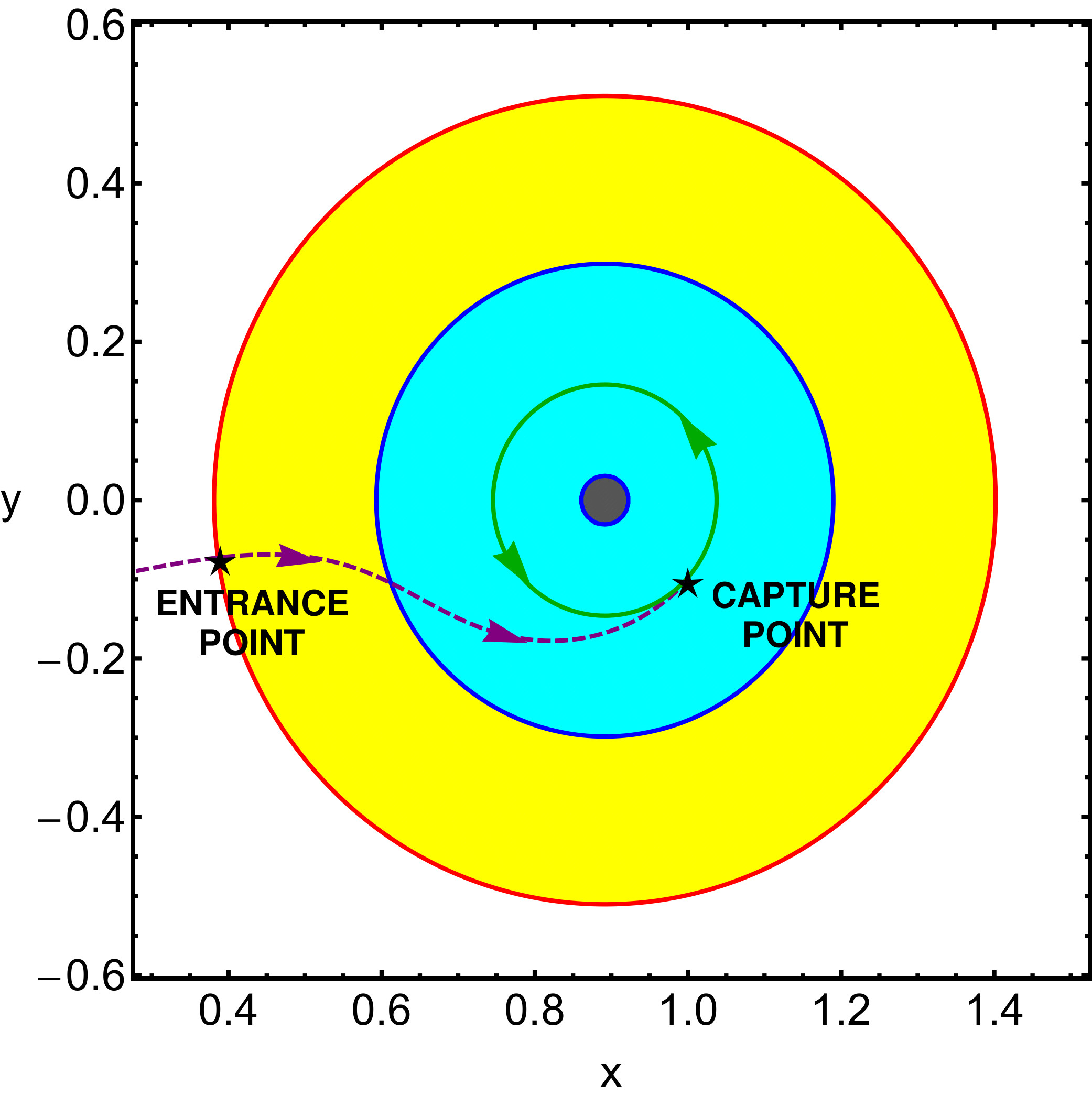}}
\caption{Schematic depicting the position of Charon (gray dot) along with the scattering region (cyan annulus) and Charon's sphere of influence (red circle). More details, regarding the radii of the circular regions, are given in the text. The pre-maneuver orbit is shown in dashed purple line, while the post-maneuver orbit is shown in solid green line. The position of the entrance and capture points are pinpointed by five-pointed black stars. (Color figure online).}
\label{disk}
\end{figure}

In this section, the gravitational capture in the pre-maneuver portion is investigated to decease the pre-maneuver velocity $V_1$. Similar to \cite{QX14}, we also assume that capture points are apsis points with respect to Charon, including periapsides and apoapsides. Since the velocity direction of the test particle at the capture point is perpendicular to the vector from the center of Charon to the test particle, the velocity direction has only two possibilities in the orbital plane: the prograde motion (anticlockwise motion with respect to Charon) or the retrograde motion (clockwise motion with respect to Charon).

The gravitational capture requires that the test particle starting from the capture point can leave the sphere of influence (SOI) of Charon under a given flight time by backward integration. The radius of the SOI of Charon is set as 10000 km (see Fig. \ref{disk}), about double the distance of $L_1$ from the center of the smaller primary, which is similar to the lunar SOI proposed by \cite{YKI92}. The flight time from the boundary of the SOI to the capture point is restricted to 15 time units (approximately 15.24 days). For the given position and velocity direction of the capture point, we postulate that a larger the velocity magnitude can make the escape from Charon (that is the Charon capture in the forward propagation) easier. Under the limitation of the flight time, we can use the following method to find the minimum $V_1$ for the given capture point.
\begin{enumerate}
	\item The initial point of backward integration is an apsis point with respect to Charon chosen from a scattering region. The scattering region of initial points is an annulus region around Charon, whose inner and outer radii are equal to the radius of Charon and the distance of $L_1$ from Charon, i.e., 606.0 and 5848.3 km, respectively (see Fig. \ref{disk}). Inside this annulus we define a uniform and dense grid of $1024 \times 1024$ initial conditions. There are two kinds of direction of velocity: the prograde and retrograde motion, but its velocity magnitude in the Pluto-Charon rotating frame $V_1$ can be changed.
	\item According to Eq. (\ref{ham}), $V_1 = \sqrt{2U(x,y) - C}$, so the minimum $V_1$ can translate to the maximum $C$. For the given capture point $(x,y)$ and Jacobi constant $C$, its velocity in the Pluto-Charon rotating frame, $(\dot{x}, \dot{y})$, can be obtained from
	\begin{equation}
	\label{vel}
	\begin{split}
	&V_1 = \sqrt{2U(x,y)-C},\\
	&\dot{x} = - k V_1 y/r_2,\\
	&\dot{y} = k V_1(x + \mu - 1)/r_2,
	\end{split}
	\end{equation}
	\noindent where $k$ denotes the direction of the velocity: $k = 1$ for the prograde motion with respect to Charon; $k = -1$ for the retrograde motion with respect to Charon. $r_2$ is the distance of the initial point to the center of Charon, and can be obtained from Eq. (\ref{dist}). In this way, the initial state of the capture point $(x,y,\dot{x},\dot{y})$ in the rotating frame can be derived.
	\item We choose the initial $C$ as $C_1$. Then, using the method in step 2, we can obtain the corresponding initial state $(x,y,\dot{x},\dot{y})$. By backward integration, a capture orbit can be obtained from the initial state. If the orbit is bounded in the SOI within 15 time units or collides with Charon, we postulate that the given $C$ cannot satisfy the gravitational capture. Then we test whether the backward escape can be achieved for a smaller $C$. In this way, $C$ is decreasing until the backward escape can be satisfied. The corresponding maximum $C$ is defined as $C_{max}$.
\end{enumerate}

It should be noted that the above numerical method is similar to that in \citep{QX14}, but in step 3, we use $C$ as the variable rather than $V_1$ in \citep{QX14}. The advantage of $C$ is that its initial value or upper boundary is specific, i.e. $C_1$.

\begin{figure*}[!t]
\centering
\resizebox{0.8\hsize}{!}{\includegraphics{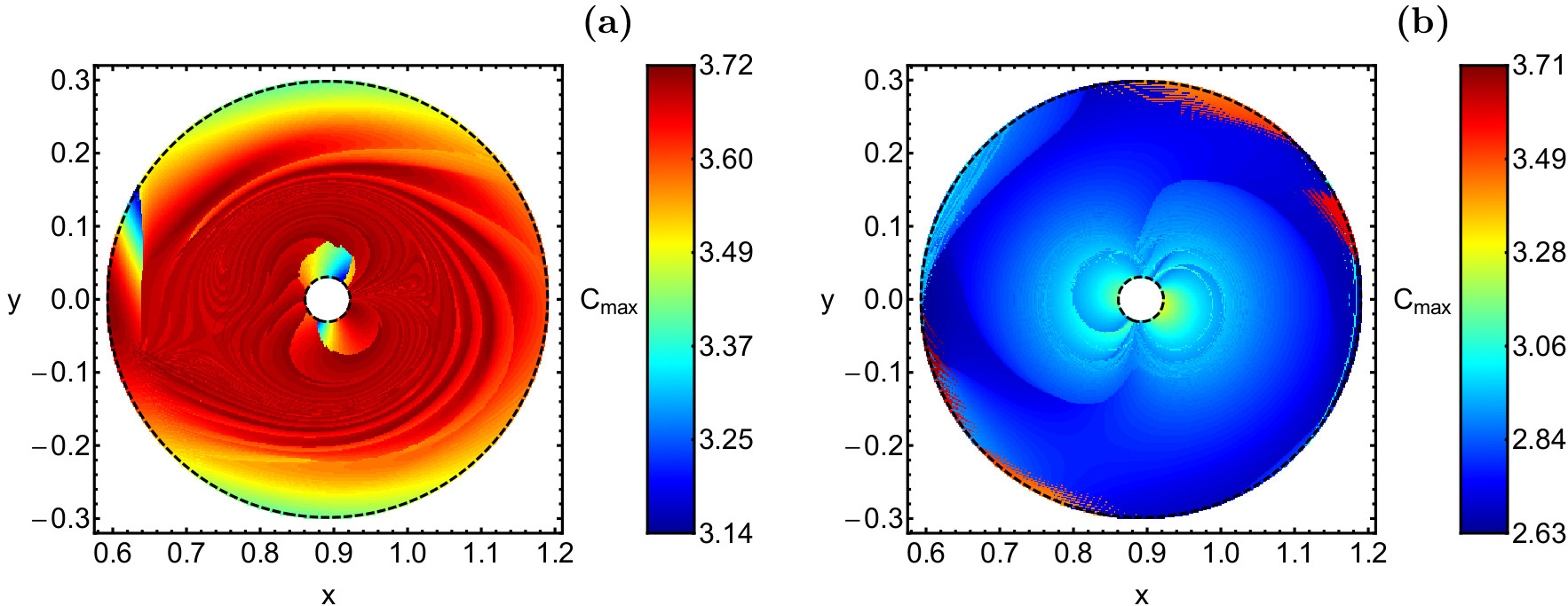}}
\caption{Distributions of $C_{max}$ in the dimensionless Pluto-Charon rotating frame. (a-left): Prograde motion and (b-right): retrograde motion. (Color figure online).}
\label{cmax}
\end{figure*}

In Fig. \ref{cmax} we display the distributions of $C_{max}$ in the Pluto-Charon rotating frame, where left and right panels correspond to prograde and retrograde capture points, respectively. From Fig. \ref{cmax}, we discover that $C_{max}$s of the prograde motion are larger than those of the retrograde motion in most areas. The lower boundary of $C_{max}$ of the retrograde motion is significantly smaller than that of prograde motion. Compared with distributions of $C_{max}$ of Earth-Moon PCRTBP in \citep{QX14}, structures of $C_{max}$ for retrograde motion are very similar. For the prograde motion, the distribution of $C_{max}$ in the Earth-Moon PCRTBP \citep{QX14} is more regular and consecutive, and there exist two branches with smaller $C_{max}$ in two sides of the Moon; however, in the Pluto-Charon PCRTBP, highly-fractal
diffused structures of $C_{max}$ are more complicated in the Charon realm.

\begin{figure*}[!t]
\centering
\resizebox{0.8\hsize}{!}{\includegraphics{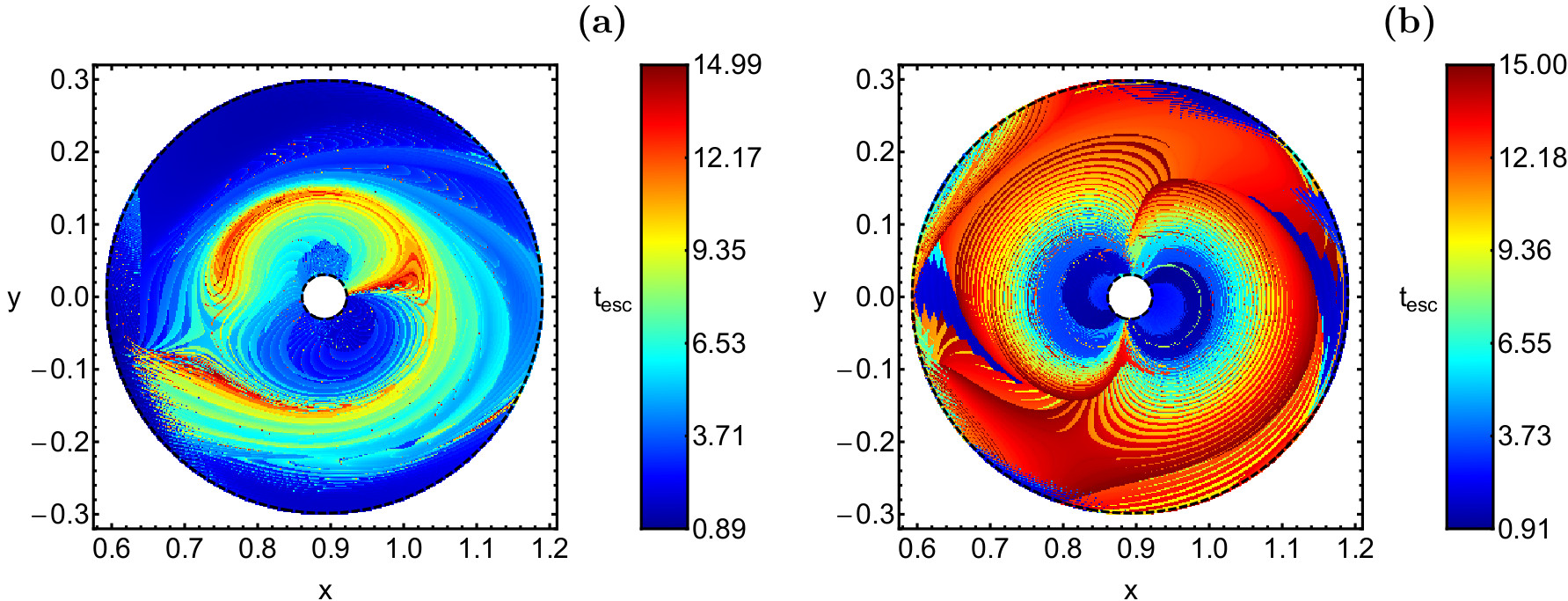}}
\caption{Distributions of $t_{esc}$ in the dimensionless Pluto-Charon rotating frame. (a-left): Prograde motion and (b-right): retrograde motion. (Color figure online).}
\label{tesc}
\end{figure*}

In the calculation of $C_{max}$, the integration time from the initial point to the boundary of SOI of Charon is denoted by $t_{esc}$. Fig. \ref{tesc} shows the distributions of $t_{esc}$ in the Pluto-Charon rotating frame. Similarly, left and right panels correspond to prograde and retrograde capture points, respectively. By looking at the figure, it can be seen that in most region, orbits starting from prograde capture points are easier than those from the retrograde points to escape the Charon's SOI by backward integration. In addition, we can find some similarities between structures of $t_{esc}$ and $C_{max}$ in Fig. \ref{cmax}.

\begin{figure*}[!t]
\centering
\resizebox{0.8\hsize}{!}{\includegraphics{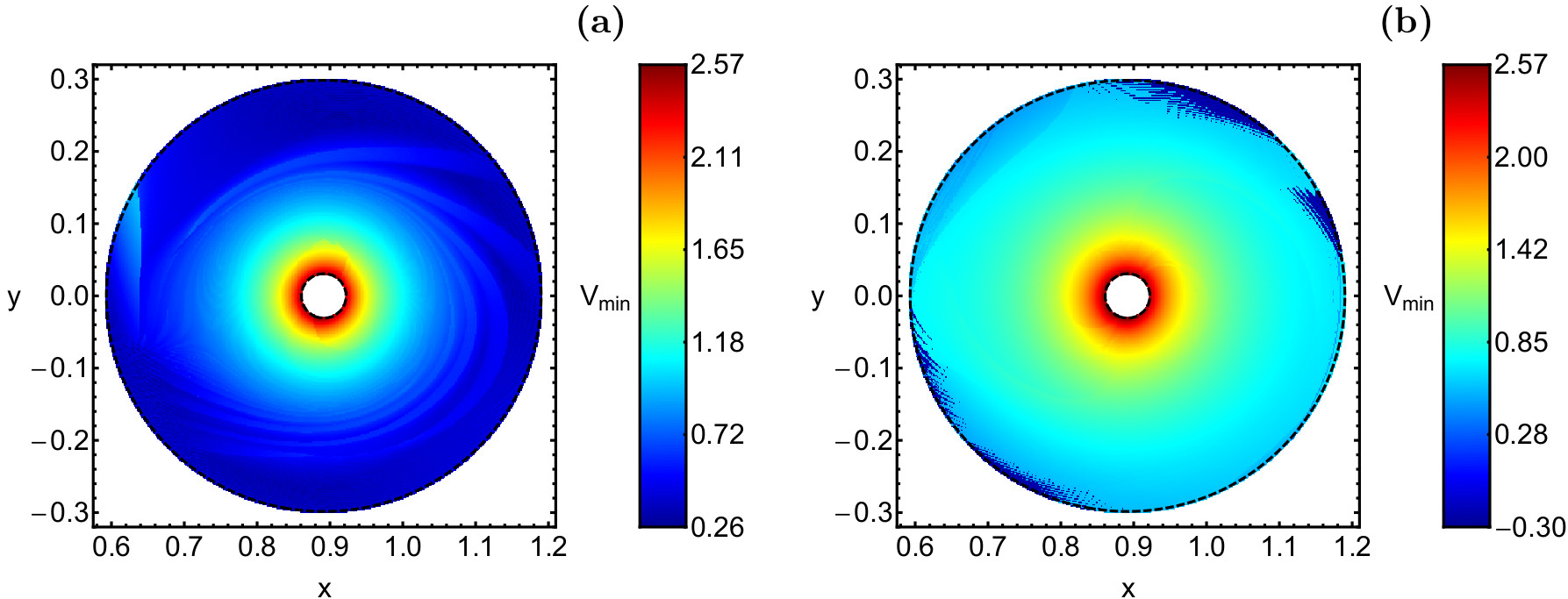}}
\caption{Distributions of $V_{min}$ in the dimensionless Pluto-Charon rotating frame. (a-left): Prograde motion and (b-right): retrograde motion. (Color figure online).}
\label{dvmin}
\end{figure*}

According to Eq. (\ref{vel}), the minimum $V_1$ we seek can be obtained by $C_{max}$. It should be noted that this minimum velocity is a magnitude in the Pluto-Charon rotating frame rather than in the Charon-centered inertial frame. We define the corresponding minimum velocity in the Charon-centered inertial frame as $V_{min}$. Since capture points are apsides with respect to Charon, we can obtain the following relationship,
\begin{equation}
\label{vmin}
V_{min} = \sqrt{2U(x,y) - C_{max}} +kr_2.
\end{equation}
Fig. \ref{dvmin} shows the distributions of $V_{min}$ in the Pluto-Charon dimensionless rotating frame. The left panel corresponds to prograde capture points, while the right one corresponds to retrograde capture points.

From Fig. \ref{dvmin}, we observe that, generally, if the capture point is farther from Charon, the minimum capture velocity $V_{min}$ is smaller. This result is accordant with the fact: the attraction from Charon at a capture point farther to Charon is smaller, which results in a smaller escape velocity. As we can see from the figure, for both prograde and retrograde motions, minimum capture velocity $V_{min}$s are mainly distributed in the range of 0.28 $\sim$ 2.57. However, in Fig. \ref{cmax} $C_{max}$ in the prograde orbits is obviously larger than that in the retrograde orbits in most areas. We can explain this phenomenon by Eq. (\ref{vmin}): for the same $V_{min}$ at a given capture point, $C_{max}$ of prograde motion is larger than that of retrograde motion. $C_{max}$ is derived from the state of the rotating frame, so the influence of the Pluto-Charon rotation can influence $C_{max}$.

In the two-body model, we can use the orbital eccentricity $e$ to determine whether the orbit is elliptic or hyperbolic. In the vicinity of a primary body of the PCRTBP, since the gravity of that celestial body is dominant, the concept of the eccentricity $e$ can still be valid to study the orbital character of the test particle. However, it should be noted that in the PCRTBP, $e$ is time-varying. When $e$ is alternated from larger than 1 to smaller than 1, we postulate that gravitational capture happens. The minimum capture eccentricity $e_{min}$ is defined as the eccentricity at the capture point corresponding to the minimum $V_{min}$, and can be obtained from
\begin{equation}
\label{emin}
e_{min} = \frac{r_2 V_{min}^2}{\mu} - 1.
\end{equation}

\begin{figure*}[!t]
\centering
\resizebox{0.8\hsize}{!}{\includegraphics{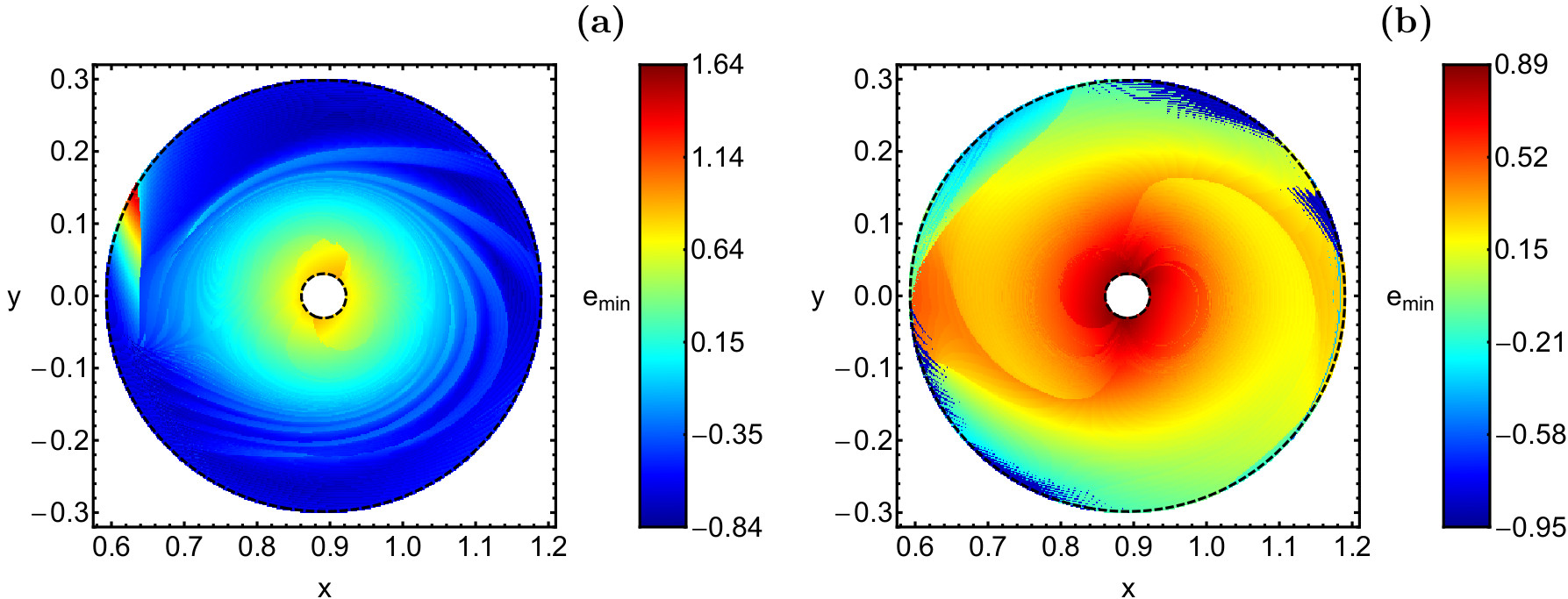}}
\caption{Distributions of $e_{min}$ in the dimensionless Pluto-Charon rotating frame. (a-left): Prograde motion and (b-right): retrograde motion. (Color figure online).}
\label{demin}
\end{figure*}

Fig. \ref{demin} illustrates distributions of the minimum capture eccentricity $e_{min}$ in the Pluto-Charon rotating frame. Based on the definition of the gravitational capture, capture points with $e_{min}>1$ are infeasible for the gravitational capture. We should avoid these regions when we choose the capture points. As we can see from the figure, in the right figure, all retrograde capture points are available, and the structures are similar to those of $C_{max}$. \cite{QX14} found that structures of $e_{min}$ of the Earth-Moon PCRTBP for prograde motion have two branches in two sides of the Moon with $e_{max}>1$. However, as we can see from the left figure, the infeasible prograde capture points for gravitational capture are only distributed in a small region on the left edge.

\section{Long-term capture}
\label{postm}

In this section, we focus on the maximum post-maneuver velocity $V_2$ satisfying the long-term capture around Charon. According to \citep{QX14}, a regular orbit in KAM tori around Charon can be bounded in the Charon realm permanently in the Pluto-Charon PCRTBP, otherwise, an orbit located in chaotic areas escapes the Charon realm. Therefore, we require that a test particle after $\Delta V$ should be inserted into a regular orbit around Charon. In \citep{QX14}, KAM tori were applied to find regular bounded regions. Although this method is simple and visible, the position and coverage of KAM tori rely on the empirical observation rather than the quantitative analysis. The SALI is a common index for quantitatively determining the regular or chaotic character of an orbit \citep{S01}. The analytical definition of SALI, as well as the particular threshold values for distinguishing between ordered and chaotic motion are given in \citet{ZJ18}. The classifications of the long-term orbits around Charon can be obtained by the value of the SALI quantitatively, including (i) bounded (regular or chaotic), (ii) escaping and (iii) collision \citep{Z15}. Essentially, other diagnostic indexes for the orbital chaos, such as the the fast Lyapunov indicator (FLI) \citep{AF04}, the correlation dimension \citep{BTG08}, and the frequency analysis method \citep{L90}, can also be applied to distinguish stable regular orbits from chaotic ones. In this section, we use the SALI to investigate bounded basins of the long-term orbits around Charon.\par

\begin{figure*}[!t]
\centering
\resizebox{0.8\hsize}{!}{\includegraphics{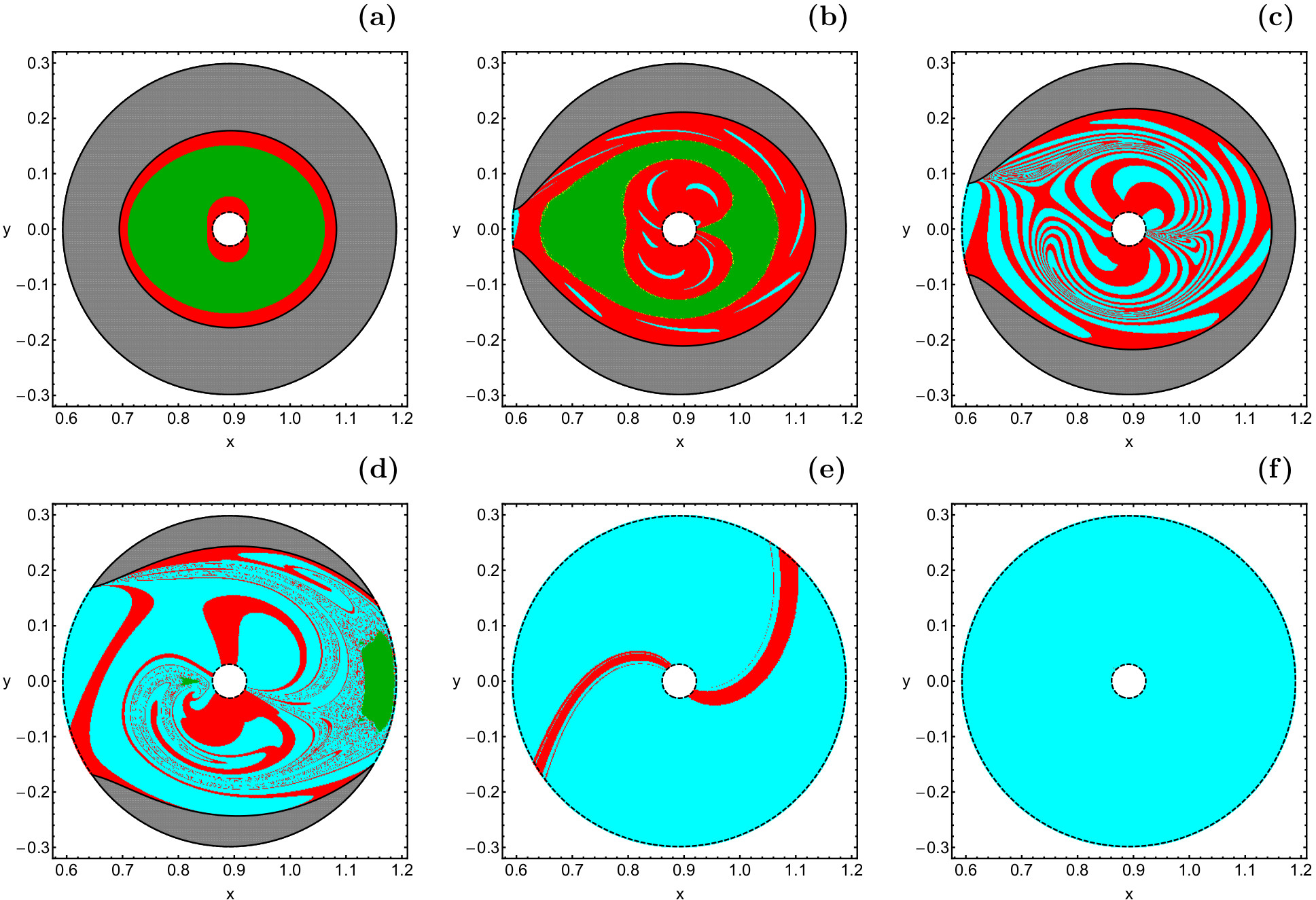}}
\caption{Basin diagrams on the configuration $(x,y)$ plane for the prograde motion, where (a): $C = 3.9$; (b): $C = 3.71$; (c): $C = 3.68$; (d): $C = C_2$; (e): $C = C_3$; (f): $C = C_4$. The color code is as follows: non-escaping regular orbits (green); trapped chaotic orbits (yellow); escaping orbits (cyan); collision orbits (red); energetically forbidden regions (gray). (Color figure online).}
\label{prg}
\end{figure*}

In Fig. \ref{prg} we present the orbital structure of the configuration $(x,y)$ plane for the prograde motion. It is seen that when $C > C_1$ all the energetically allowed area of motion is covered either by collision or by regular initial conditions of orbits. As long as $C < C_1$ the channel around $L_1$ opens and therefore orbits can escape to the Pluto realm. The stable areas of regular initial conditions are destroyed abruptly. It should be noted that when $C = C_2$ two new stability islands emerge at both sides of Charon. We observe that as the value of the Jacobi constant decreases the amount of both regular and collision orbits decrease. On the other hand, the amount of escaping orbits grows rapidly and for $C \leq C_4$ they dominate the entire configuration plane. In the Earth-Moon PCRTBP \citep{QX14}, KAM tori rapidly disappear when $C$ decreases to $C_2$, but apparently, regular orbits in the Pluto-Charon system can still survive when $C=C_2$. We postulate that this difference is derived from the larger $\mu$ of Pluto-Charon system, because orbits rotating around the smaller primary with a larger $\mu$ have a greater resistance against the perturbation from the larger primary in the PCRTBP.

When $C < C_1$, the abrupt emergence of chaotic regions in Fig. \ref{prg} looks quite similar to the abrupt bifurcation to fully developed chaotic scattering in \citep{BLE90}. However, we postulate that the chaotic phenomenon in Fig. 7 essentially is different from the chaotic scattering. Firstly, the fully developed chaotic scattering denotes a situation in which all periodic orbits are unstable and there are no KAM surfaces (i.e., the dynamics is hyperbolic) \citep{BLE89}, but in Fig. 7, obviously, there exist KAM tori in the bounded regular region (denoted by green) when $C$ is smaller than $ C_1$ \citep{QX14}. In addition, in the CRTBP we consider, not all periodic orbits are unstable, for example the distant retrograde orbits around the smaller primary are stable \citep{HEN69,LAM05}. Secondly, \cite{K00} pointed out that the chaotic phenomenon in the region around the smaller primary of the CRTBP is ascribed to the existence of transversal homoclinic and heteroclinic points. We can explain these chaotic phenomena by the symbolic dynamics or the horseshoe-like dynamics \citep{K00}.

It should be noted that trapped chaotic motion is very limited. This means that the corresponding initial conditions appear only as lonely and isolated points, mainly in the vicinity of the stability islands of regular motion. Indeed, in panel (b) of Fig. \ref{prg} one can distinguish several near the boundaries of the bounded basins of regular motion the presence of several initial conditions which correspond to trapped chaotic motion. 

\begin{figure*}[!t]
\centering
\resizebox{0.8\hsize}{!}{\includegraphics{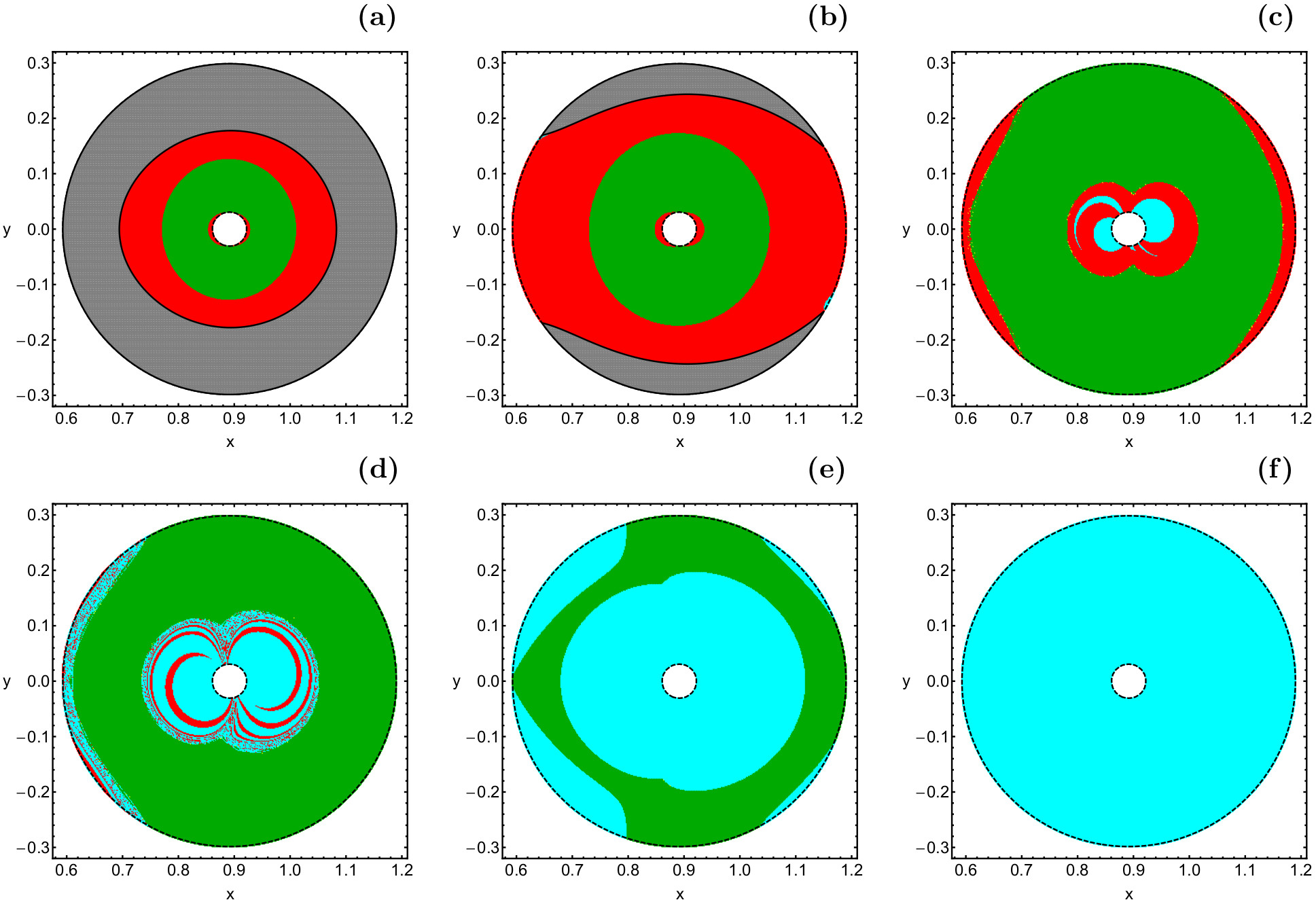}}
\caption{Basin diagrams on the configuration $(x,y)$ plane for the retrograde motion, where (a): $C = 3.9$; (b): $C = C_2$; (c): $C = C_4$; (d): $C = 2.9$; (e): $C = 2.8$; (f): $C = 2.5$. The color code is as follows: non-escaping regular orbits (green); trapped chaotic orbits (yellow); escaping orbits (cyan); collision orbits (red); energetically forbidden regions (gray). (Color figure online).}
\label{rtg}
\end{figure*}

In the same vein, the orbital structure of the configuration plane for the retrograde motion is depicted in Fig. \ref{rtg}. It is evident that the orbital structure corresponding to the retrograde motion has many differences with respect to that of the prograde motion. The most notable ones are: (i) bounded regular motion is possible for much lower values of the Jacobi constant, or alternatively, for higher values of the total orbital energy; (ii) the overall orbital structure seems less complicated, without the presence of regions with highly fractal basin boundaries. Those results are similar to those of the Earth-Moon PCRTBP in\citep{QX14}.

\begin{figure*}[!t]
\centering
\resizebox{0.8\hsize}{!}{\includegraphics{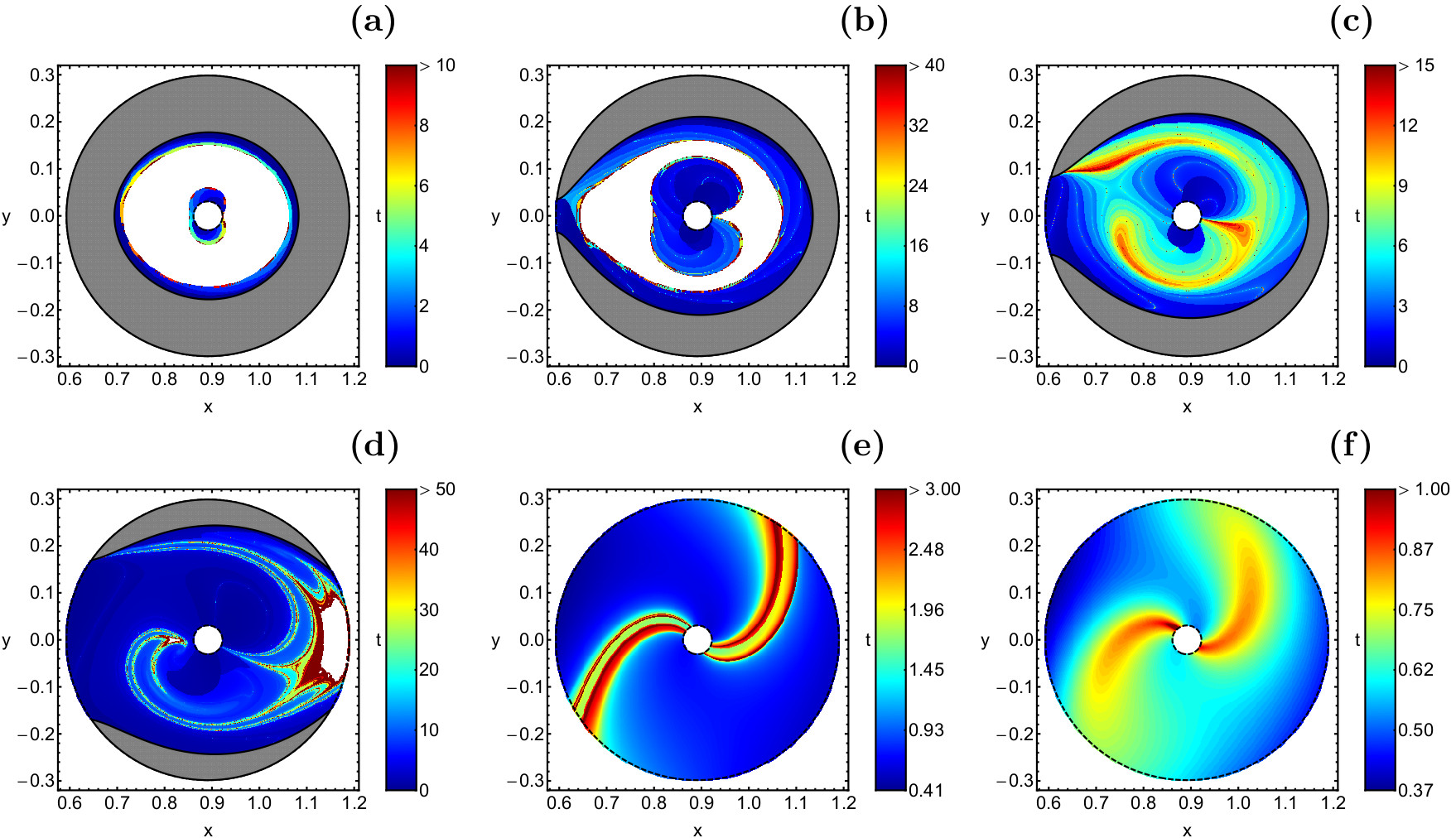}}
\caption{Distribution of the escape and collision times of the orbits for the prograde motion, where (a): $C = 3.9$; (b): $C = 3.71$; (c): $C = 3.68$; (d): $C = C_2$; (e): $C = C_3$; (f): $C = C_4$. All initial conditions of bounded (regular and chaotic) orbits are shown in white. (Color figure online).}
\label{prgt}
\end{figure*}

\begin{figure*}[!t]
\centering
\resizebox{0.8\hsize}{!}{\includegraphics{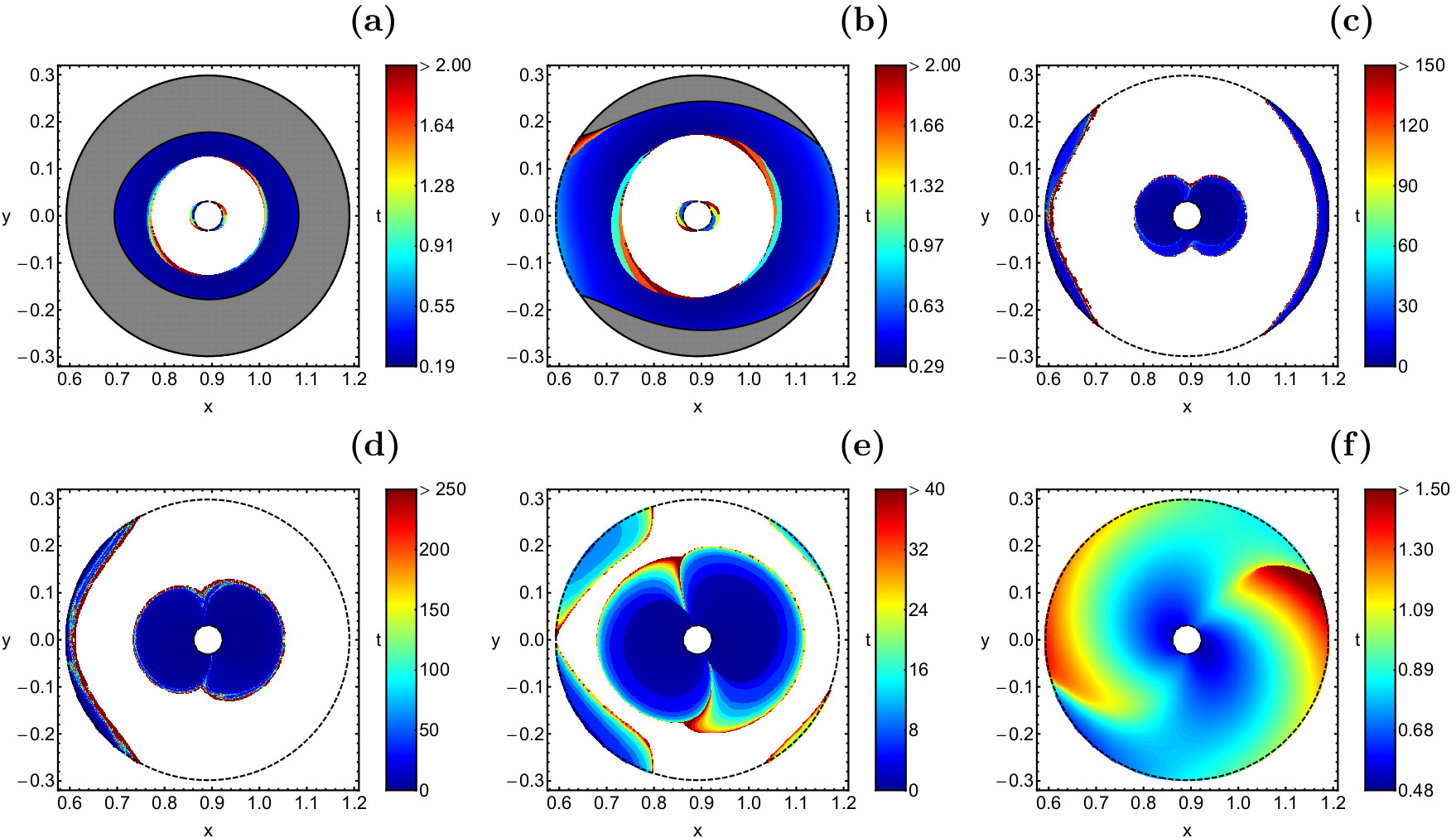}}
\caption{Distribution of the escape and collision times of the orbits for the retrograde motion, where (a): $C = 3.9$; (b): $C = C_2$; (c): $C = C_4$; (d): $C = 2.9$; (e): $C = 2.8$; (f): $C = 2.5$. All initial conditions of bounded (regular and chaotic) orbits are shown in white. (Color figure online).}
\label{rtgt}
\end{figure*}

The distributions of the corresponding escape and collision time of the orbits of both the prograde and retrograde motion are given in Figs. \ref{prgt} and \ref{rtgt}, respectively. As expected, the lowest values of both escape and collision time were measured inside the corresponding basins of escape and collision, while the highest times were identified mainly in the fractal basin boundaries, where the degree of the unpredictability of the system maximizes.

The color-coded diagrams of Figs. \ref{prg} and \ref{rtg} reveal the orbital structure of the configuration $(x,y)$ plane for a fixed value of the Jacobi constant (or equivalently, for a fixed orbital energy), for both prograde and retrograde motion. In an attempt to obtain the same information for a continuous interval of $C$ values, we define a plane of representation in which the $x$-coordinate is the abscissa, while the ordinate is given by the Jacobi constant. This means that all orbits start with initial position $(x_0, 0) $, i.e, on the $x$-axis, with initial velocities $(0,\dot{y}_{0})$, where the $y$-component of the velocity is derived from the third of Eqs. (\ref{vel}).

\begin{figure*}[!t]
\centering
\resizebox{0.8\hsize}{!}{\includegraphics{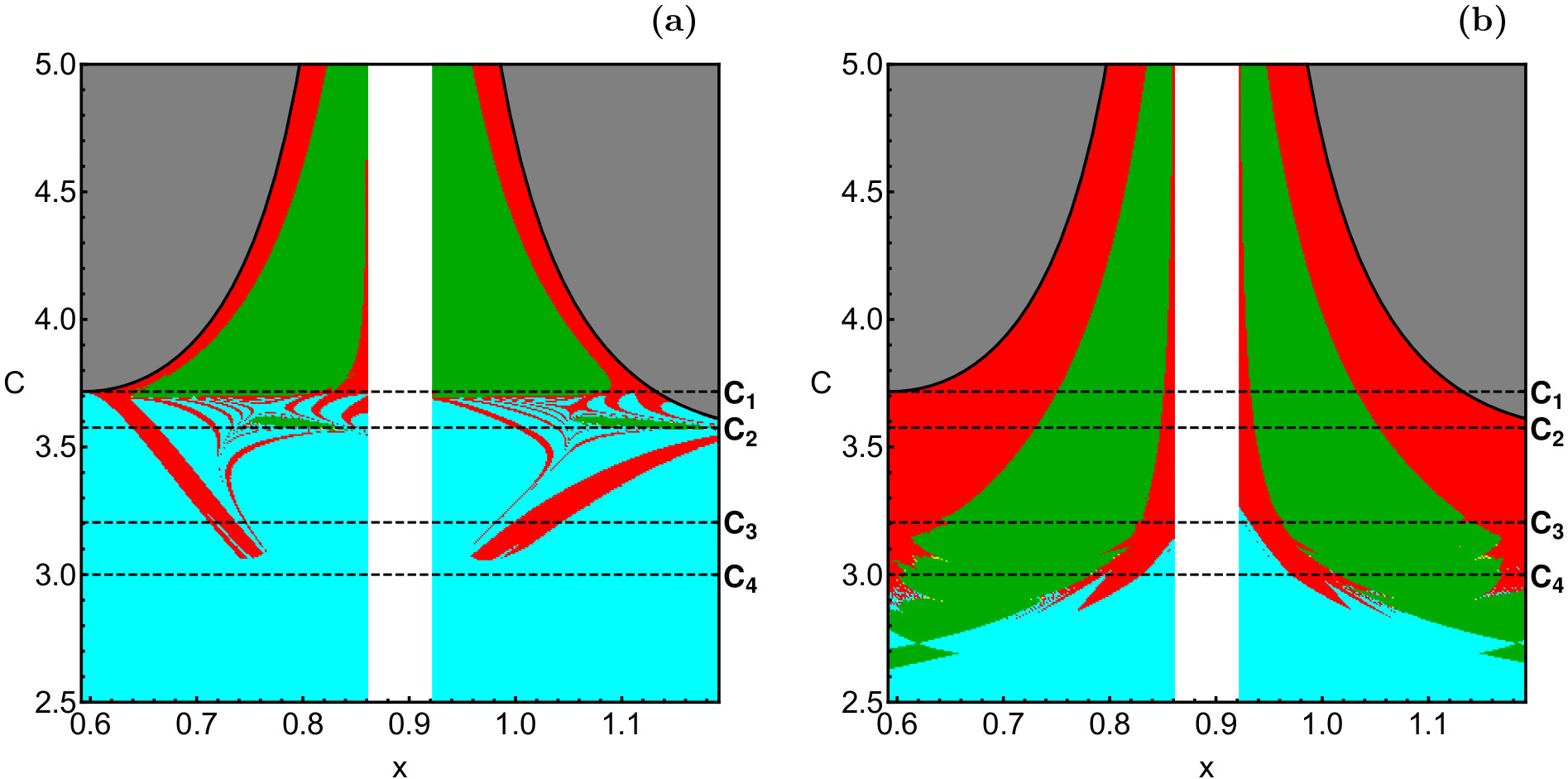}}
\caption{Basin diagrams on the $(x,C)$-plane for (a-left): the prograde motion and (b-right): the retrograde motion. The color code is as follows: non-escaping regular orbits (green); trapped chaotic orbits (yellow); escaping orbits (cyan); collision orbits (red); energetically forbidden regions (gray). (Color figure online).}
\label{xc}
\end{figure*}

\begin{figure*}[!t]
\centering
\resizebox{0.9\hsize}{!}{\includegraphics{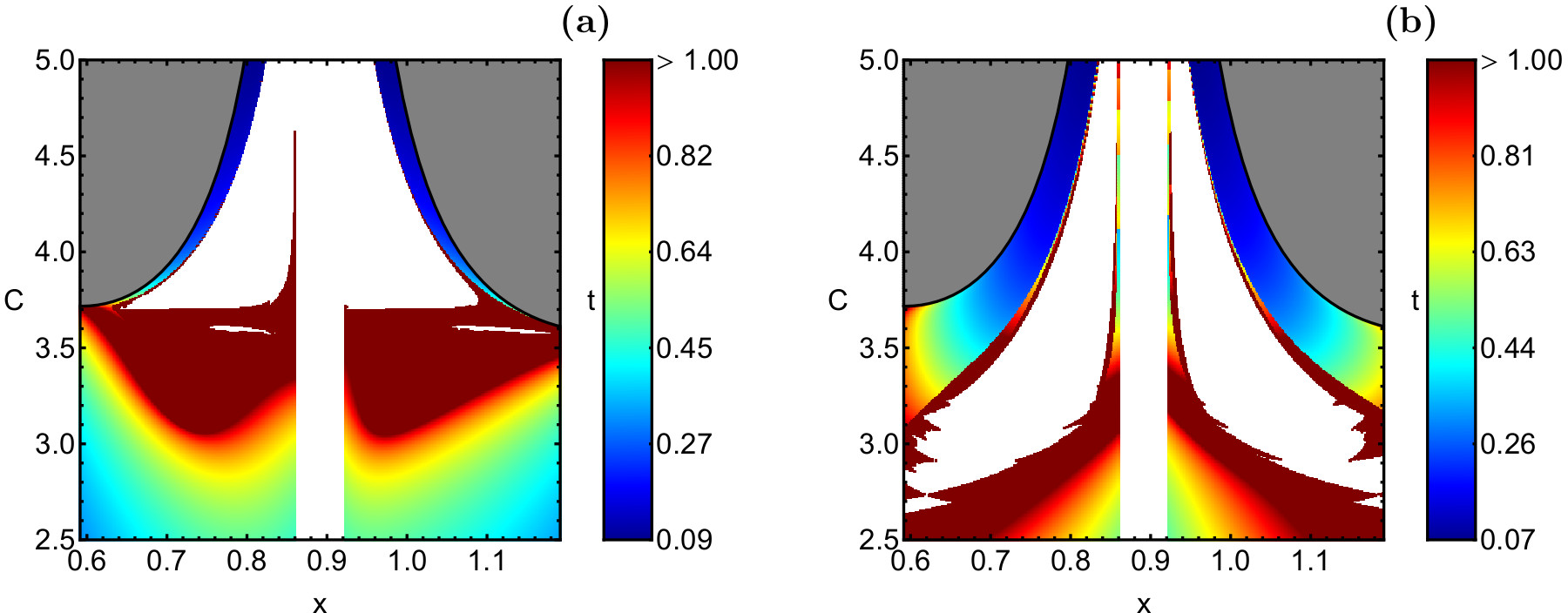}}
\caption{Distribution of the corresponding escape and collision times of the orbits for (a-left): the prograde motion and (b-right): the retrograde motion. All initial conditions of bounded (regular and chaotic) orbits are shown in white. (Color figure online).}
\label{xct}
\end{figure*}

In Fig. \ref{xc} we present the basin diagram on the $(x,C)$ plane for $C \in [2.5, 5]$, for both the prograde and retrograde motions. The black continuous line indicates the corresponding Zero Velocity Curve (ZVC), and it is defined as
\begin{equation}
f(x,C) = 2U(x,y = 0) = C.
\label{zvc}
\end{equation}
We see that for the prograde motion bounded regular orbits exist only when $C > C_2$, while for the retrograde motion the same type of orbits is still possible even below $C = C_4$. However, for extremely high and low values of $C$ we encounter the same orbital structure for both types of motion (prograde and retrograde). In particular, for extremely low values of the Jacobi constant escaping orbits completely dominate, while for relatively high values of $C$ bounded and collision types of orbits share the energetically allowed phase space. In Fig. \ref{xct} we present the distribution of times employed by the test particle in the escape and collision orbits.

Based on the previous discussion, for both prograde and retrograde motions, bounded orbits generally erode or disappear with the decrease of $C$ (see Figs \ref{prg} and \ref{rtg}). Hence, for a given capture point $(x,y)$ and its direction of motion, there exists a minimum $C$ satisfying long-term capture in the Charon region. According to Eq. (\ref{ham}), $V_2$ corresponding to the minimum $C$ can be obtained, which is the maximum $V_2$ we seek for.\par

\section{Near-optimal capture of Charon}
\label{nearo}

\begin{figure*}[!t]
\centering
\resizebox{0.8\hsize}{!}{\includegraphics{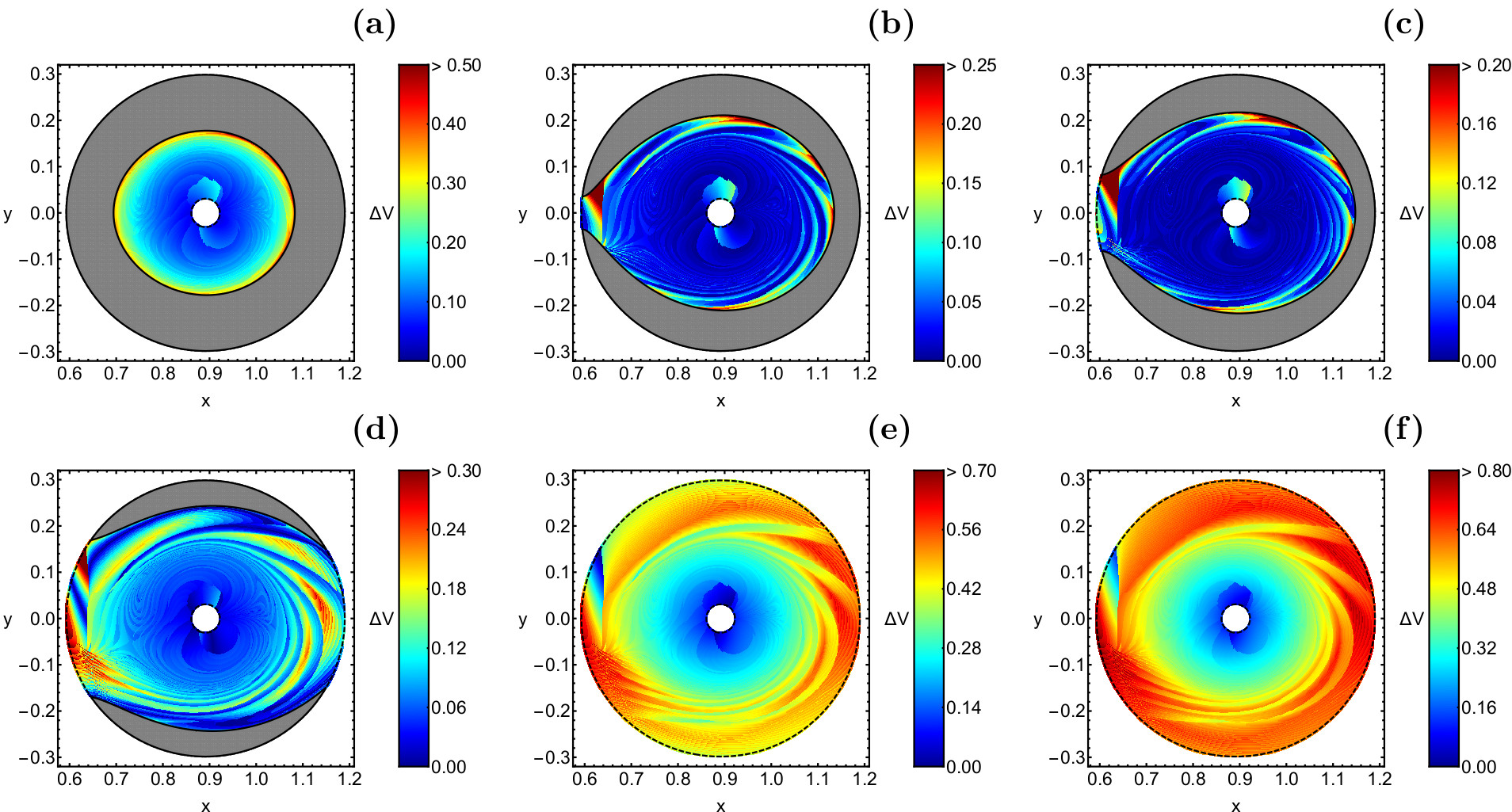}}
\caption{Distributions of $\Delta V$ for the prograde motion, where (a): $C = 3.9$; (b): $C = 3.71$; (c): $C = 3.68$; (d): $C = C_2$; (e): $C = C_3$; (f): $C = C_4$. (Color figure online).}
\label{prgv}
\end{figure*}

\begin{figure*}[!t]
\centering
\resizebox{0.8\hsize}{!}{\includegraphics{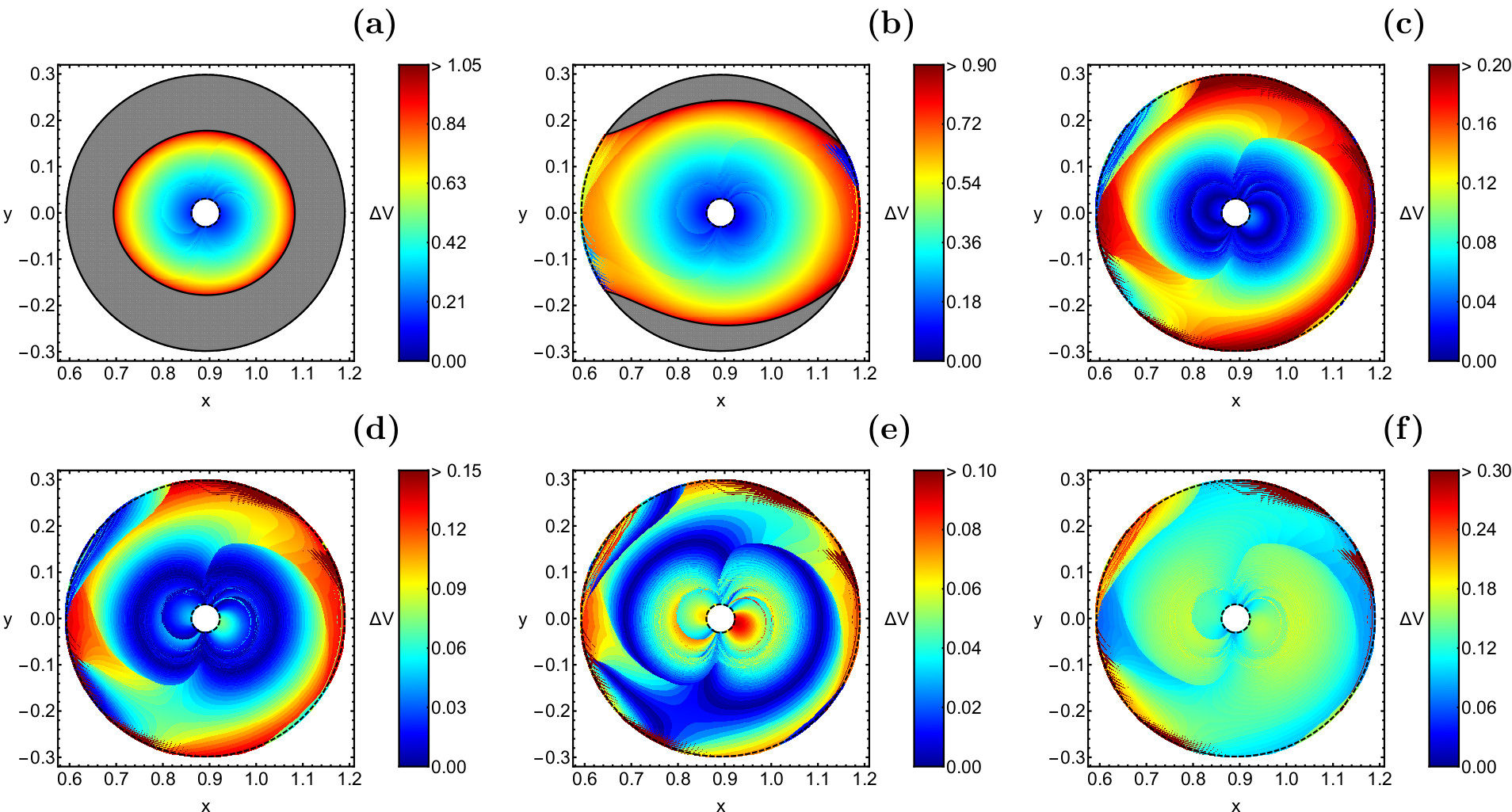}}
\caption{Distributions of $\Delta V$ for the retrograde motion, where (a): $C = 3.9$; (b): $C = C_2$; (c): $C = C_4$; (d): $C = 2.9$; (e): $C = 2.8$; (f): $C = 2.5$. (Color figure online).}
\label{rtgv}
\end{figure*}

\begin{figure*}[!t]
\centering
\resizebox{0.7\hsize}{!}{\includegraphics{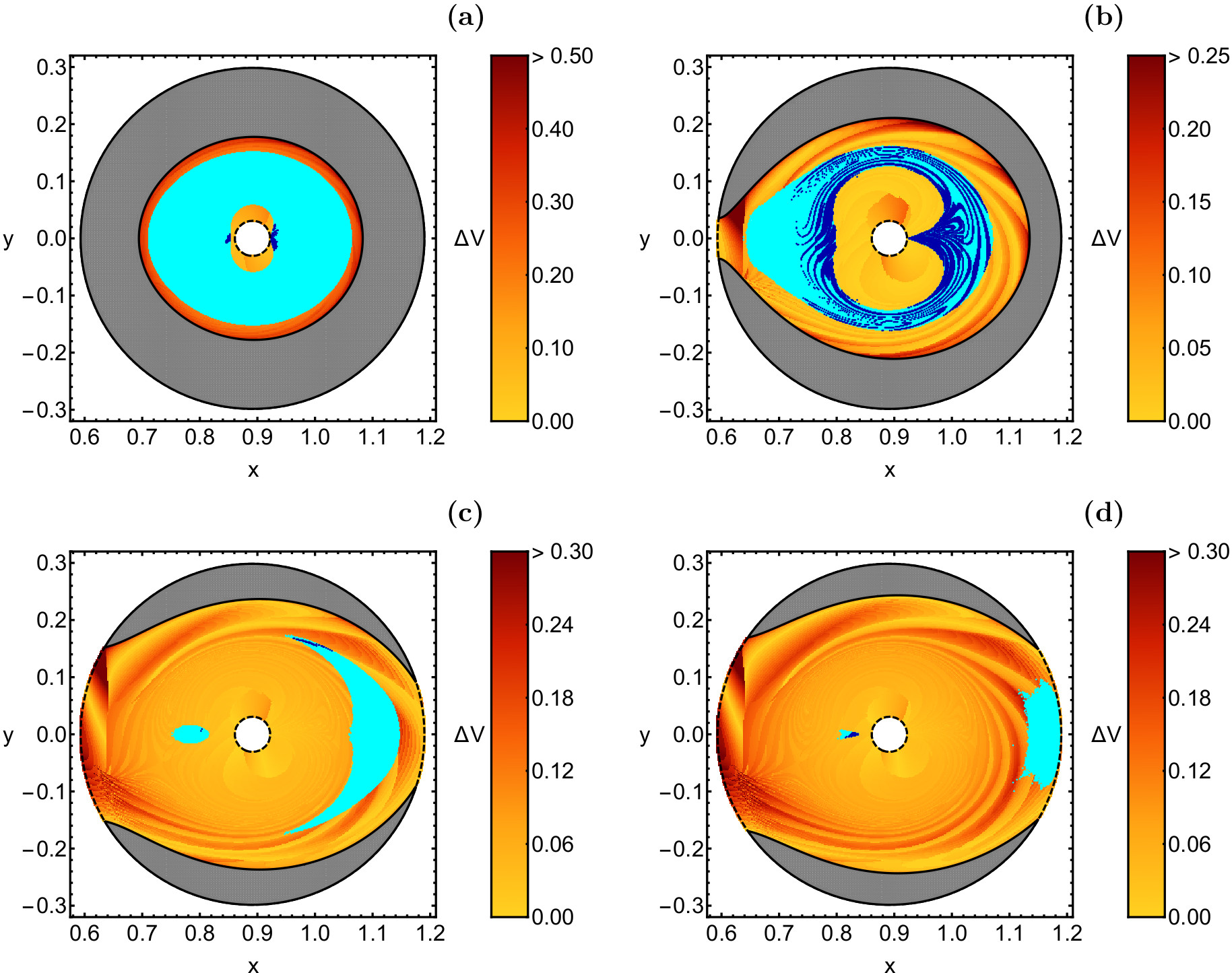}}
\caption{Overlays of $\Delta V$ and bounded regular basins, for the prograde motion, where (a): $C = 3.9$; (b): $C = 3.71$; (c): $C = 3.60$; (d): $C = C_2$. All bounded basins are indicated with cyan color, while the initial conditions of both escape and collision orbits are shown in tones of solar colors, according to the value of $\Delta V$. All near-optimal capture initial conditions are pinpointed by blue dots. (Color figure online).}
\label{prgm}
\end{figure*}

\begin{figure*}[!t]
\centering
\resizebox{0.7\hsize}{!}{\includegraphics{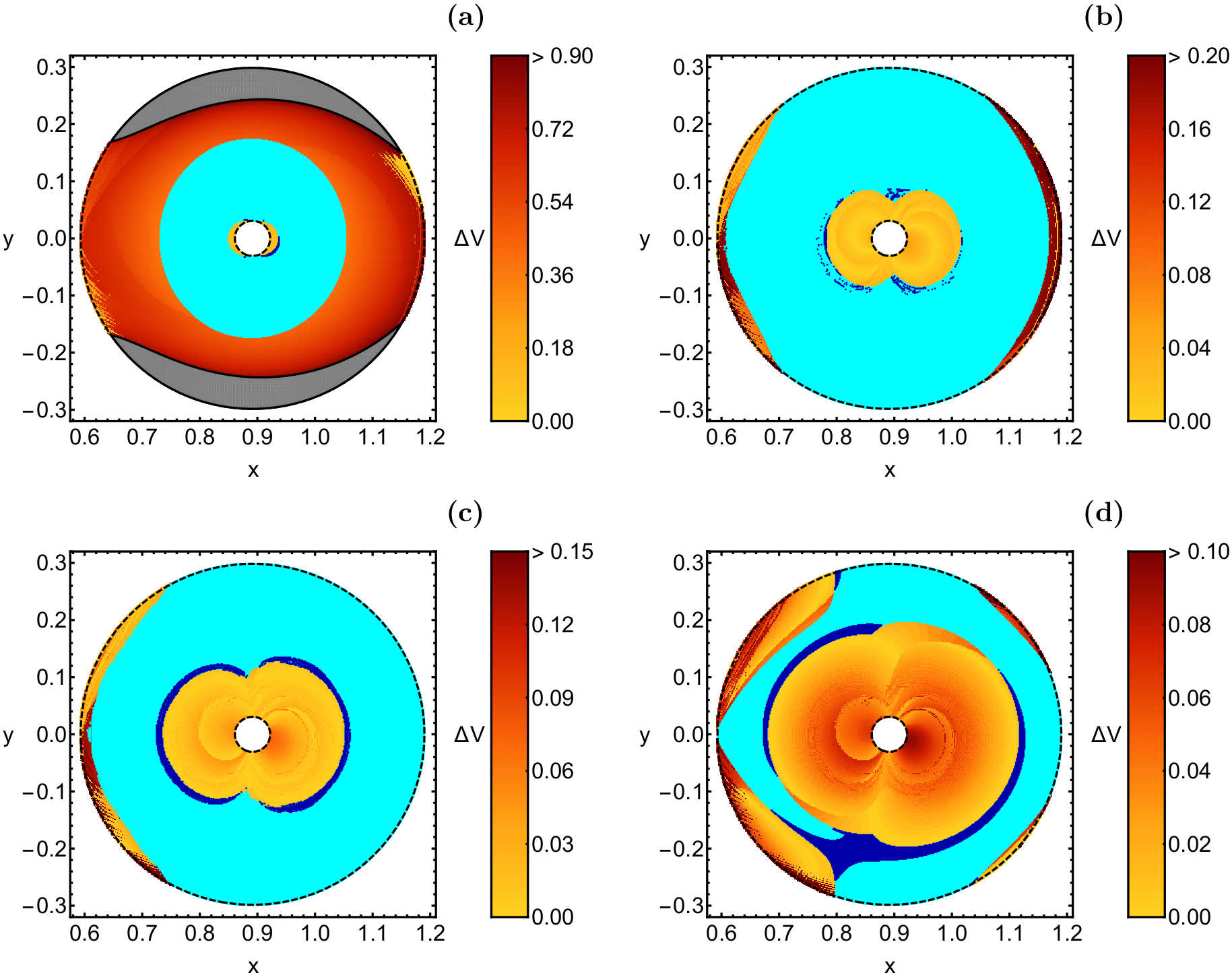}}
\caption{Overlays of $\Delta V$ and bounded regular basins, for retrograde motion, where (a): $C = C_2$; (b): $C = C_4$; (c): $C = 2.9$; (d): $C = 2.8$. All bounded basins are indicated with cyan color, while the initial conditions of both escape and collision orbits are shown in tones of solar colors, according to the value of $\Delta V$. All near-optimal capture initial conditions are pinpointed by blue dots. (Color figure online).}
\label{rtgm}
\end{figure*}

In the previous two sections, we have investigated how to obtain the minimum $V_1$ and the maximum $V_2$, successively. In this section, we synthesize those results to analyse the minimum $\Delta V$. According to \cite{QX14}, the key to find the minimum $\Delta V$ is to locate the near-optimal capture point. Combining results obtained in previous sections, near-optimal capture points can be found by the following method:

\begin{enumerate}
	\item For the given velocity direction, $\Delta V$ corresponding to the capture point $(x,y) $ can be calculated by
	\begin{equation}
	\label{deltav}
    \begin{split}
    \Delta V &= |V_1 - V_2| \\
    &= |\sqrt{2U(x,y) - C_{max}} - \sqrt{2U(x,y) - C}|
    \end{split}	
	\end{equation}
	\noindent where $C_{max}$ is obtained from Figs. \ref{cmax} or \ref{tesc}, and $C$ is the Jacobi constant of the post-maneuver portion. It should be noted that if the given capture point $(x,y) $ is located in the energetically forbidden regions for the given $C$, Eq. (\ref{deltav}) is invalid and the corresponding $\Delta V$ does not exist.
	\item Using Eq. (\ref{deltav}), the distribution of $\Delta V$ in the scattering region for given velocity direction and $C$ can be obtained. For example, Figs. \ref{prgv} and \ref{rtgv} display the distributions of $\Delta V$ in the Pluto-Charon rotating frame corresponding to the cases presented in Figs. \ref{prg} and \ref{rtg}, respectively. In addition, for the same velocity direction and $C$, the corresponding basin diagrams can be obtained using the method in Section \ref{postm} (see Figs. \ref{prg} and \ref{rtg}). Then, we put the bounded basins onto the distribution of $\Delta V$, and get an overlay figure of them. For example, Figs. \ref{prgm} and \ref{rtgm} display overlays for different $C$s and motions. The background of the overlay is the distribution of $\Delta V$, and cyan regions denote bounded basins.
	\item From the overlay figures, points in bounded basins denote feasible points in terms of the long-term capture after $\Delta V$. Among these feasible capture points, those located in the minimum $\Delta V$ regions are chosen as near-optimal capture points. For example, blue points in Figs. \ref{prgm} and \ref{rtgm} are near-optimal capture points we chose. The corresponding $\Delta V$ is the minimum $\Delta V$ we seek for.
\end{enumerate}

Figs. \ref{prgv} and \ref{rtgv} display the distributions of $\Delta V$ corresponding to Figs. \ref{prg} and \ref{rtg}, respectively. Gary regions in those figures are the energetically forbidden regions in the Charon realm for larger $C$s. Compared Fig. \ref{prgv} with Fig. \ref{cmax} (a), the structure of the distribution of $\Delta V$ for the prograde motion actually is quite similar to that of $C_{max}$ for prograde motion. It can be explained by Eq. (\ref{deltav}). Similarly, the structure of the distribution of $\Delta V$ in Fig. \ref{rtgv} is similar to that of $C_{max}$ for retrograde motion in Fig. \ref{cmax} (b).

Figs. \ref{prgm} and \ref{rtgm} show the overlays of $\Delta V$ and the bounded basins of different $C$ for the prograde and retrograde motions, respectively. Gray and cyan regions in these figures denote energetically forbidden regions and bounded basins, respectively. Dark blue points are near-optimal capture points we chose. Based on the analysis of Fig. \ref{prg}, only for a larger $C$ can bounded basins of the prograde motion exist. Hence, near-optimal prograde capture points will disappear with the decrease of $C$. But near-optimal retrograde capture points can survive in smaller $C$s. By looking at these figures, it can be seen that since distributions of $\Delta V$ and bounded basins of prograde motion are both quite different from those of retrograde motion, the distributions of the corresponding near-optimal capture points have significant differences between two kinds of motions. For example, in Fig. \ref{rtgm}, near-optimal retrograde capture points are mostly located in the inner boundary of bounded basins near Charon, and similar to the distribution of bounded basins, near-optimal capture points gradually move away from Charon with the decrease of $C$. However, situations of near-optimal prograde capture points are more complicated according to Fig. \ref{prgm}, where regions with highly fractal basin boundaries exist, especially when $C = 3.71$. In Fig. \ref{prgm} (a), near-optimal capture points are distributed in the inner boundary of bounded basin nearest Charon. But when $C$ decreases to 3.71, i.e., necks around $L_1$ just open, near-optimal capture points are distributed in a highly fractal pattern around Charon, which is similar to the structures near Charon in Figs. \ref{cmax} and \ref{prgv}. When $C$ decreases to 3.60, corresponding bounded basins are distributed in two separated stable islands, we can obtain two near-optimal capture regions in two separated bounded basins, respectively. But when $C$ decreases to $C_2$, near-optimal capture points can only be distributed in a small region of the left island. Numerical computation indicates that minimum $\Delta V$s corresponding to near-optimal capture points in Figs. \ref{prgm} and \ref{rtgm} are all smaller than 23 m/s. \par

\begin{figure*}[!t]
\centering
\resizebox{0.6\hsize}{!}{\includegraphics{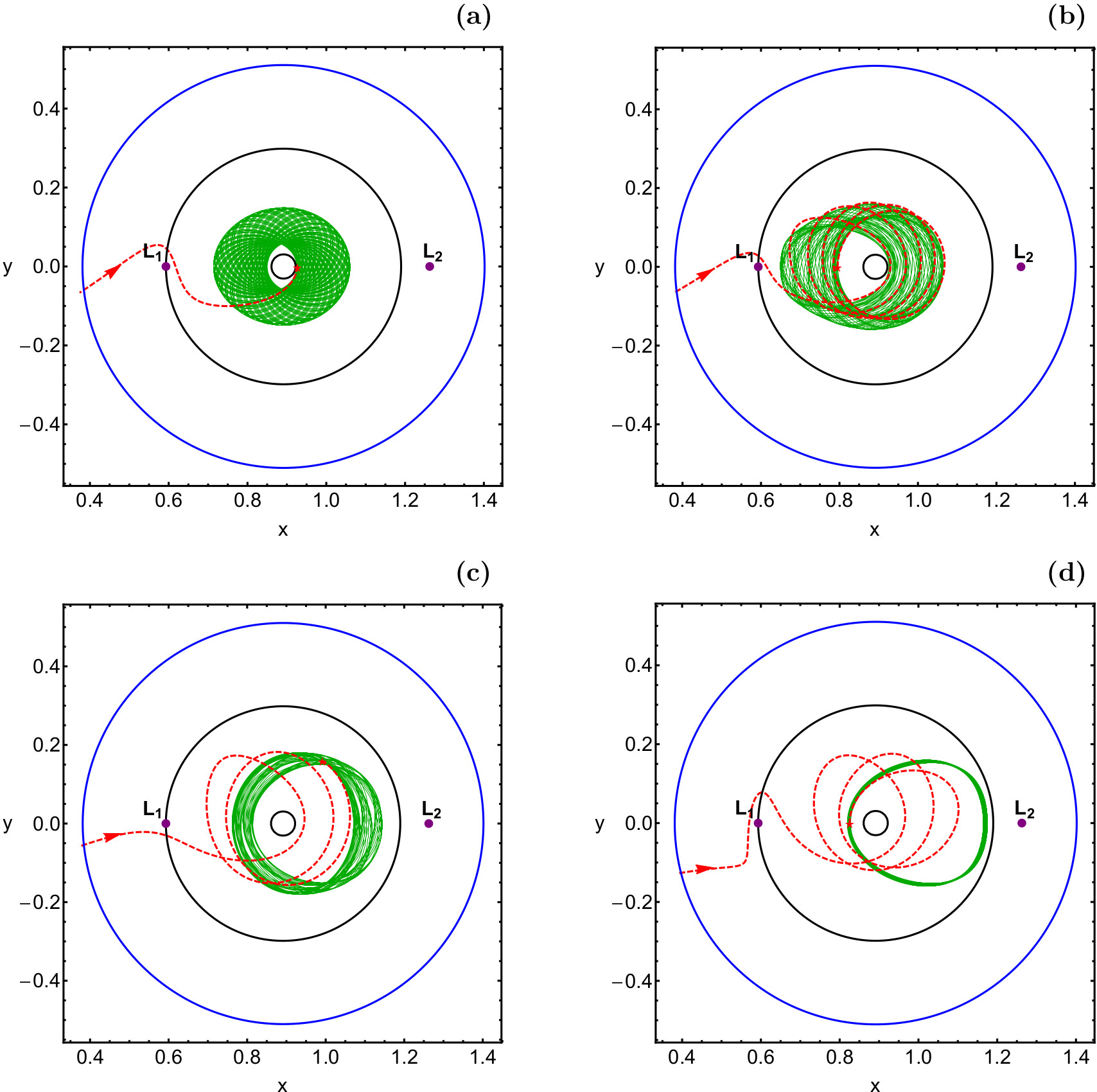}}
\caption{Capture obits corresponding to the prograde near-optimal capture points, where the Jacobi constants of the post-maneuver orbit are (a): $C = 3.9$; (b): $C = 3.71$; (c): $C = 3.60$; (d): $C = C_2$. The pre-maneuver orbits are shown in dashed, red dashed lines, while the post-maneuver orbits are shown in solid green lines. The black solid lines indicate the boundaries of the scattering region, while the blue solid line delimits the SOI of Charon. The position of the Lagrange points $L_1$ and $L_2$ are pinpointed by purple dots. (Color figure online).}
\label{orbsp}
\end{figure*}

\begin{figure*}[!t]
\centering
\resizebox{0.6\hsize}{!}{\includegraphics{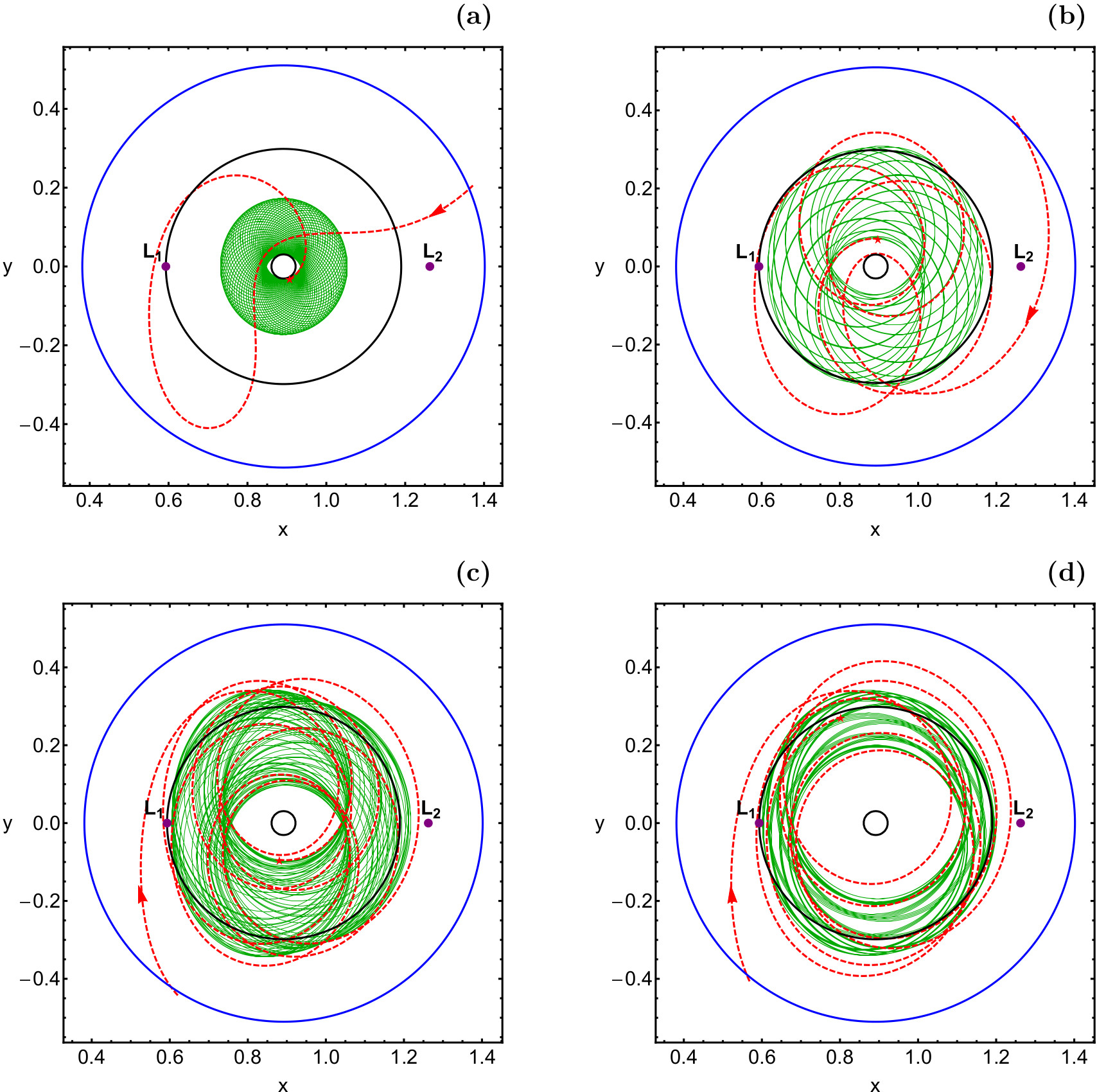}}
\caption{Capture obits corresponding to the retrograde near-optimal capture points, where the Jacobi constants of the post-maneuver orbit are (a): $C = C_2$; (b): $C = C_4$; (c): $C = 2.9$; (d): $C = 2.8$. The pre-maneuver orbits are shown in dashed, red dashed lines, while the post-maneuver orbits are shown in solid green lines. The black solid lines indicate the boundaries of the scattering region, while the blue solid line delimits the SOI of Charon. The position of the Lagrange points $L_1$ and $L_2$ are pinpointed by purple dots. (Color figure online).}
\label{orbsr}
\end{figure*}

Using the near-optimal capture points as the initial state, the corresponding pre- and post-maneuver capture orbits can be constructed. In Figs. \ref{orbsp} and \ref{orbsr}, we illustrate the capture orbits corresponding to the near-optimal capture points in Figs. \ref{prgm} and \ref{rtgm}, respectively. The figures are depicted in the dimensionless Pluto-Charon rotating frame. Red dashed lines denote pre-maneuver portions of capture orbits, while green solid lines denote post-maneuver portions of capture orbits (or long-term capture orbits). Red star points are the optimal capture points (or impulsive points) chosen from the near-optimal capture regions in Figs. \ref{prgm} and \ref{rtgm} correspondingly. By looking at the figures, we observe that both pre- and post-maneuver portions of capture orbits have significant differences between prograde and retrograde motions. In Fig. \ref{orbsp}, since the corresponding $C$s of prograde motion are larger than $C_2$, the sizes of post-maneuver orbits are smaller than those of retrograde motion in Fig. \ref{orbsr}. In addition, all pre-maneuver portions of prograde near-optimal capture points pass through the gateway of $L_1$. However, in Fig. \ref{orbsr}, most pre-maneuver portions of retrograde near-optimal capture points pass through the gateway of $L_2$. As we can see Fig. \ref{orbsr}, multiple Charon flybys occur during the pre-maneuver portions. Data of prograde and retrograde near-optimal capture orbits in Figs. \ref{orbsp} and \ref{orbsr} are listed in Tables \ref{table1} and \ref{table2}. From these tables, $\Delta V$s in Fig. \ref{orbsp} are smaller than 11 m/s, while $\Delta V$s in Fig. \ref{orbsr} are smaller than 23 m/s. Of particular note is that based on the data of Fig. \ref{orbsp} (c) and (d), $C_{max}$s of pre-maneuver portions are larger than the corresponding $C$s of post-maneuver portions, so the impulsive maneuvers at the near-optimal capture points increase the speed instead of brake.

\begin{table}
	\centering
	\caption{Data of prograde near-optimal capture orbits.}
	\label{table1}
    \setlength{\tabcolsep}{1pt}
	\begin{tabular}{lcccr}
		\hline
		Fig. \ref{orbsp} & $(x,y)$ & $C_{max}$ & $C$ & $\Delta V$ (m/s) \\ \hline
		(a) & $(0.92575062,-0.00083565)$ & 3.687 &  3.90 & 10.4064 \\
		(b) & $(0.79538850, 0.00083565)$ & 3.697 &  3.71 &  1.2940 \\
        (c) & $(0.99093168, 0.15961003)$ & 3.627 &  3.60 &  6.0325 \\
        (d) & $(0.82547207,-0.00083565)$ & 3.677 & $C_2$ &  7.3500 \\
		\hline
	\end{tabular}
\end{table}

\begin{table}
	\centering
	\caption{Data of retrograde near-optimal capture orbits.}
	\label{table2}
    \setlength{\tabcolsep}{1pt}
	\begin{tabular}{lcccr}
		\hline
		Fig. \ref{orbsr} & $(x,y)$ & $C_{max}$ & $C$ & $\Delta V$ (m/s) \\ \hline
		(a) & $(0.90736622,-0.03091922)$ & 3.097 & $C_2$ & 22.6293 \\
        (b) & $(0.89733836, 0.07103064)$ & 2.947 & $C_4$ &  3.5683 \\
        (c) & $(0.88062527,-0.09442897)$ & 2.867 &  2.90 &  2.5561 \\
        (d) & $(0.80207374, 0.27158774)$ & 2.777 &  2.80 &  3.1574 \\
		\hline
	\end{tabular}
\end{table}
The above results we obtained, including optimal capture points and corresponding capture orbits, could be applied in future space mission design (like the New Horizons interplanetary space probe). Although in real space missions there are several types of plane perturbations mainly due to the presence of other giant celestial bodies, such as Jupiter and Saturn, all these perturbations are extremely weak in the vicinity of Charon (which was the scattering region in our work). Therefore, we argue that the numerical results of the Pluto-Charon CRTBP obtained in this paper are structurally stable against out of plane perturbations.

\section{Conclusions}
\label{conc}

In this paper, the near-optimal capture problem in the Pluto-Charon system was investigated and discussed. In an attempt to achieve long-term capture around Charon with the minimum $\Delta V$, optimal capture points were found and located by the near-optimal capture theory. According to the definition of $\Delta V$, the capture orbit was divided into the pre- and post-maneuver portions. In the pre-maneuver portion, the maximum escape Jacobi constant $C_{max}$ was obtained and discussed by backward numerical integration. In the post-maneuver portion, the SALI was applied to numerically investigate the orbital characters of long-term capture orbits around Charon. The initial conditions corresponding to three types of motion: (1) bounded, (2) escaping and (3) collisional were classified and distinguished. Combining the results of pre- and post-maneuver portions, optimal capture points were located by overlay figures. Furthermore, corresponding capture orbits were constructed in the PCRTBP. In addition, the near-optimal capture results in the Pluto-Charon system of this paper were compared with those in the Earth-Moon system. We postulated that the significant difference of mass ratio $\mu$ resulted in the different results between two planet-moon systems.

The main results of our numerical research can be summarized as follows:
\begin{enumerate}
\item Numerical computation indicated that the $C_{max}$s of the prograde motion are larger than those of the retrograde motion in most areas. Compared with distributions of $C_{max}$ of the Earth-Moon PCRTBP in \citep{QX14}, the structures of $C_{max}$ for prograde motion in the Pluto-Charon PCRTBP are more complicated in the Charon realm, where highly-fractal diffused structures of $C_{max}$ exist. Based on the results of $t_{esc}$, in most region, orbits starting from prograde capture points are easier than those from retrograde points to escape the Charon's SOI by backward integration.
\item Based on the definition of gravitational capture and distributions of $e_{min}$, we found that all retrograde capture points are available for gravitational capture, and its structure of $e_{min}$ is similar to that of $C_{max}$. Different from the two branches in two sides of the Moon for prograde motion in the Earth-Moon PCRTBP \citep{QX14}, the prograde infeasible capture points for gravitational capture in the Pluto-Charon PCRTBP are only distributed in a small region on the left edge.
\item According to the orbital structure of long-term orbits around Charon, we found that the orbital structure corresponding to the retrograde motion has obvious differences with respect to that of the prograde motion: (1) bounded regular motion is possible for much lower values of the Jacobi constant, or alternatively, for higher values of the total orbital energy; (2) the overall orbital structure seems less complicated, without the presence of regions with highly fractal basin boundaries. In addition, the lowest values of both escape and collision time were measured inside the corresponding basins of escape and collision, while the highest times were identified mainly in the fractal basin boundaries, where the degree of the unpredictability of the system maximizes.
\item The distributions of near-optimal capture points indicated that near-optimal prograde capture points disappear rapidly with the decrease of $C$, but near-optimal retrograde capture points can survive in smaller $C$s. Near-optimal retrograde capture points are mostly located in the inner boundary of bounded basins near Charon, and gradually move away from Charon with the decrease of $C$. However, situations of prograde motion are more complicated, where highly fractal patterns around Charon and separated stable islands exist.
\item Numerical results indicated that all pre-maneuver portions of prograde near-optimal capture points pass through the gateway of $L_1$. However, most pre-maneuver portions of retrograde near-optimal capture points pass through the gateway of $L_2$. Data of both prograde and retrograde near-optimal capture orbits showed that, $\Delta V$s are smaller than 23 m/s.
\end{enumerate}

For the numerical integration (both backward and forward in time) of the equations of motion (\ref{eqmot}), as well as of the variational equations (\ref{variac}) we used a double precision Bulirsch-Stoer \verb!FORTRAN 77! algorithm \citep{PTVF92}, with a fixed time step equal to $10^{-3}$. In all our calculations the numerical error, related to the conservation of the Jacobi integral (\ref{ham}) was generally smaller than $10^{-12}$, while in most of the cases it was smaller than $10^{-14}$. All the graphical illustration has been created using the latest version 11.2 of the software Mathematica$^{\circledR}$ \citep{W03}.

The obtained results in this paper, including optimal capture points and corresponding capture orbits, could be applied in future space mission design.

\section*{Acknowledgments}

Yi Qi wishes to acknowledge the support of the Natural Sciences and Engineering Research Council of Canada through a Discovery Accelerator Supplement under Grant RGPAS-493042-2016. The authors would like to express their warmest thanks to the two anonymous reviewers for the careful reading of the manuscript as well as for all the valuable suggestions and comments which allowed us to improve both the quality and the clarity of the paper.


\begin{thebibliography}{}

\bibitem[\protect\citeauthoryear{Astakhov et al.}{2003}]{ABW03} Astakhov, S.A., Burbanks, A.D., Wiggins, S., Farrelly, D., 2003. Chaos-assisted capture of irregular moons. Nature 423, 264-267.

\bibitem[\protect\citeauthoryear{Astakhov \& Farrelly}{2004}]{AF04} Astakhov, S.A., Farrelly, D., 2004. Capture and escape in the elliptic restricted three-body problem. Monthly Notices of the Royal Astronomical Society 354, 971-979.

\bibitem[\protect\citeauthoryear{Bao et al.}{2015}]{BYB15} Bao, C., Yang, H., Barsbold, B., Baoyin, H., 2015. Capturing near-Earth asteroids into bounded Earth orbits using gravity assist. Astrophysics and Space Science 360, 61.

\bibitem[\protect\citeauthoryear{Baoyin et al.}{2010}]{BCL10} Baoyin, H-.X., Chen, Y., Li, J.-F., 2010. Capturing near earth objects. Research in Astronomy and Astrophysics 10, 587-598.

\bibitem[\protect\citeauthoryear{Belbruno}{1987}]{B87} Belbruno, E.A., AIAA, DGLR, and JSASS, International Electric Propulsion Conference, 19th, Colorado Springs, CO, May 11-13, 1987.

\bibitem[\protect\citeauthoryear{Belbruno \& Miller}{1990}]{BM90} Belbruno, E.A., Miller, J., Jet Propulsion Lab., Pasadena, CA, 1990.

\bibitem[\protect\citeauthoryear{Belbruno et al.}{2008}]{BTG08} Belbruno, E., Topputo, F., Gidea, M., 2008. Resonance transitions associated to weak capture in the restricted three-body problem. Advances in Space Research 42, 1330-1351.

\bibitem[\protect\citeauthoryear{Bleher et al.}{1990}]{BLE90} Bleher, S., Grebogi, C., Ott, E., 1990. Bifurcation to chaotic scattering. Physica D: Nonlinear Phenomena 46(1), 87-121.

\bibitem[\protect\citeauthoryear{Bleher et al.}{1989}]{BLE89} Bleher, S., Ott, E., Grebogi, C., 1989. Routes to chaotic scattering. Physical review letters 63(9), 919.

\bibitem[\protect\citeauthoryear{Buie et al.}{2012}]{BTG12} Buie, M.W., Tholen, D.J., Grundy, W.M., 2012. The Orbit of Charon is Circular. AJ 144, article id. 15.

\bibitem[\protect\citeauthoryear{Granvik et al.}{2012}]{GVJ12} Granvik, M., Vaubaillon, J., Jedicke, R., 2012. The population of natural Earth satellites. Icarus 218, 262-277.

\bibitem[\protect\citeauthoryear{Henon}{1969}]{HEN69} Hénon, M., 1969. Numerical exploration of the restricted problem, V. Astronomy and Astrophysics 1, 223-238.

\bibitem[\protect\citeauthoryear{Jehn et al.}{2004}]{JCG04} Jehn, R., Campagnola, S., Garcia, D., Kemble, S., 18th International Symposium on Space Flight Dynamics, 2004.

\bibitem[\protect\citeauthoryear{Koon et al.}{2000}]{K00} Koon W S, Lo M W, Marsden J E, et al., Dynamical systems, the three-body problem and space mission design, World Scientific, 2000.

\bibitem[\protect\citeauthoryear{Lam \& Whiffen}{2005}]{LAM05} Lam, T., Whiffen G J., AAS/AIAA Spaceflight Mechanics Conference, 2005, paper AAS 05-110.

\bibitem[\protect\citeauthoryear{Laskar}{1990}]{L90} Laskar, J., 1990. The chaotic motion of the solar system: A numerical estimate of the size of the chaotic zones. Icarus 88, 266-291.

\bibitem[\protect\citeauthoryear{NASA Space Science Data Coordinated Archive}{Update: 23 December 2016}]{url} NASA Space Science Data Coordinated Archive: \url{http://nssdc.gsfc.nasa.gov/planetary/factsheet/plutofact.html}

\bibitem[\protect\citeauthoryear{Press et al.}{1992}]{PTVF92} Press H.P., Teukolsky S.A, Vetterling W.T., Flannery B.P., Numerical Recipes in FORTRAN 77, 2nd Ed., Cambridge Univ. Press, Cambridge, USA, 1992.

\bibitem[\protect\citeauthoryear{Qi \& de Ruiter}{2018}]{Qd18} Qi, Y., de Ruiter, A., 2018. Short-term capture of the Earth-Moon system. Monthly Notices of the Royal Astronomical Society 476 (4), 5464-5478.

\bibitem[\protect\citeauthoryear{Qi \& Xu}{2014}]{QX14} Qi, Y., Xu, S., 2014. Lunar capture in the planar restricted three-body problem. Celestial Mechanics and Dynamical Astronomy 120, 401-422.

\bibitem[\protect\citeauthoryear{Qi \& Xu}{2016}]{QX16} Qi, Y., Xu, S., 2016. Earth--Moon transfer with near-optimal lunar capture in the restricted four-body problem. Aerospace Science and Technology 55, 282-291.

\bibitem[\protect\citeauthoryear{Roncoli \& Fujii}{2016}]{RF10} Roncoli, R.B., Fujii, K.K., AIAA Meeting Papers on Disc, 2010.

\bibitem[\protect\citeauthoryear{Schoenmaekers et al.}{2001}]{SHP01} Schoenmaekers, J., Horas, D., Pulido, J.A., 16th International Symposium on Space Flight Dynamics, Pasadena, California, 2001.

\bibitem[\protect\citeauthoryear{Skokos}{2001}]{S01} Skokos C., 2001. Alignment indices: a new, simple method for determining the ordered or chaotic nature of orbits.Journal of Physics A 34, 10029-10043.

\bibitem[\protect\citeauthoryear{Strange et al.}{2013}]{SLMc13} Strange, N., Landau, D., McElrath, T., Lantoine, G., Lam, T., Overview of mission design for NASA asteroid redirect robotic mission concept, Pasadena, CA: Jet Propulsion Laboratory, National Aeronautics and Space Administration, 2013.

\bibitem[\protect\citeauthoryear{Szebehely}{1967}]{S67} Szebehely V., Theory of Orbits, Academic Press, New York, 1967.

\bibitem[\protect\citeauthoryear{Topputo \& Belbruno}{2015}]{TB15} Topputo, F., Belbruno, E., 2015. Earth--Mars transfers with ballistic capture. Celestial Mechanics and Dynamical Astronomy 121, 329-346.

\bibitem[\protect\citeauthoryear{Verrier \& McInnes}{2014}]{VMc14} Verrier, P.E., McInnes, C.R., 2014. Low-Energy Capture of Asteroids onto Kolmogorov--Arnold--Moser Tori. Journal of Guidance, Control, and Dynamics 38, 330-335.

\bibitem[\protect\citeauthoryear{Wolfram}{2003}]{W03} Wolfram S., The Mathematica Book, Wolfram Media, Champaign, 2003.

\bibitem[\protect\citeauthoryear{Yamakawa et al.}{1992}]{YKI92} Yamakawa, H., Kawaguchi, J., Ishii, N., Matsuo, H., A numerical study of gravitational capture orbit in the Earth-Moon system, Spaceflight mechanics, 1992, pp. 1113--1132.

\bibitem[\protect\citeauthoryear{Yarnoz et al.}{2013}]{YSMc13} Yarnoz, D.G., Sanchez, J.P., McInnes C.R., 2013. Easily retrievable objects among the NEO population. Celestial Mechanics and Dynamical Astronomy 116, 367-388.

\bibitem[\protect\citeauthoryear{Zotos}{2015}]{Z15} Zotos, E.E., 2015. Orbit classification in the planar circular Pluto-Charon system. Astrophysics and Space Science 360, article id. 7.

\bibitem[\protect\citeauthoryear{Zotos \& Jung}{2018}]{ZJ18} Zotos, E.E., Jung, Ch., 2018. Correlating the escape dynamics and the role of the normally hyperbolic invariant manifolds in a binary system of dwarf spheroidal galaxies. International Journal of Non-Linear Mechanics 99, 182-203.

\end{thebibliography}
\end{document}